\documentclass[a4paper,11pt]{article}
\usepackage{JHEP-related/jheppub} 
\usepackage{lineno}

\usepackage{xcolor}
\usepackage{graphicx}
\usepackage{physics}
\usepackage{cancel}

\usepackage{enumitem}
\usepackage{hyperref}

\title{\boldmath Algorithmic Construction of Vector-Valued Modular Forms and Application to an $A_4\times\mathbb{Z}_4$-invariant Model of Leptons
}

\author{Q. Bonnefoy,} 
\author{M. Rausch de Traubenberg}
\author{and V. Saulquin} 
\affiliation{Universit\'e de Strasbourg, CNRS, IPHC UMR7178, F-67037 Strasbourg Cedex, France}

\emailAdd{quentin.bonnefoy@iphc.cnrs.fr}
\emailAdd{michel.rausch@iphc.cnrs.fr}
\emailAdd{victor.saulquin@iphc.cnrs.fr}

\abstract{We present an algorithmic procedure that constructs the Vector-Valued Modular Forms (VVMFs) of finite modular groups, as well as its implementation in a companion code. The usual method consists in building the first few VVMFs, whose tensor products then generate the others. We address the difficulty of systematically finding the minimal set of VVMFs that suffices to produce all higher-weight ones, and we propose a method to identify it that makes use of the Burnside-Brauer theorem. For any given group, whose basic group-theoretic and representation-theoretic description is provided by the user, the code determines the minimal set of VVMFs, explicitly builds it using general results of the theory of such functions, then performs the tensor products that generate any other VVMF. Importantly, it automatically takes into account all caveats that might arise during the study of new groups.  Our approach reproduces the known results for the finite modular groups explored in the recent phenomenological literature on modular flavor symmetries, and works for additional groups which we tested. We illustrate our algorithm and the working of our code by studying a model of leptons based on the unexplored finite modular group $A_4\times\mathbb{Z}_4$.}

\begin{document}
\maketitle
\flushbottom

\section{Introduction}

Modular symmetries arise naturally in a variety of physical contexts, such as extra-dimensio\-nal, supersymmetric or conformal field theories, string theories, black hole physics, flavor physics or the study of topological phases of matter. Their precise application to a given physical system often requires a fruitful communication and interplay with mathematics. In this paper, we focus on the modular group $\Gamma=SL(2,\mathbb{Z})$ interpreted as a global symmetry of supersymmetric field theories, whose application to flavor physics has recently been the subject of intense scrutiny (see e.g. \cite{Feruglio:2019ybq,Ding:2023htn,Feruglio:2025ztj} for reviews). Although we focus on the modular group $\Gamma$, other choices are interesting and actively studied \cite{deMedeirosVarzielas:2019cyj,Liu:2020msy,Yao:2020zml,Ding:2020zxw,Carducci:2026shs}.

Field theories are usually defined by a field content and a Lagrangian. Often, the latter is taken to be the most general one invariant under a given symmetry group, whose action on the fields is postulated. For $\Gamma$-invariant supersymmetric theories, there exists a complex scalar field $\tau$ known as the modulus, on which $\Gamma$ acts non-linearly \cite{Feruglio:2017spp}. The modulus is often understood as describing dynamical extra dimensions of space \cite{Ferrara:1989bc,Ferrara:1989qb,Giveon:1994fu}, in which case modular symmetry corresponds to the gauge symmetry that connects different values of $\tau$ that describe the same underlying geometry, and thus the same physics, but the mathematics of modular-invariant field theories can be formulated without reference to this specific paradigm \cite{Feruglio:2017spp}. In order to define the theory, one simply ought to specify how modular symmetry acts on the various fields, following the rules which we review in Section~\ref{sec:modInvFT}. In particular, the transformations of the fields other than the modulus, dubbed matter fields, are characterized by an integer known as the weight (see \cite{Liu:2020msy,Yao:2020zml} for possible extensions to non-integral weights) and an irreducible representation of a finite quotient group $G_d$ of $\Gamma$. The past flavor physics literature covers the cases $G_d=\Gamma_2$ \cite{Kobayashi:2018wkl}, $\Gamma_3^{(\prime)}$ \cite{Feruglio:2017spp,Criado:2018thu,Ding:2019zxk,Yao:2020qyy,Okada:2020brs,Liu:2019khw}, $\Gamma_4^{(\prime)}$ \cite{Penedo:2018nmg,Novichkov:2018ovf,Novichkov:2019sqv,Novichkov:2020eep,Liu:2020akv}, $\Gamma_5^{(\prime)}$ \cite{Novichkov:2018nkm,Ding:2019xna,Wang:2020lxk,Wang:2021mkw}, $\Gamma_6^{(\prime)}$ \cite{Li:2021buv}, $\Gamma_7^{(\prime)}$ \cite{Ding:2020msi}, $2D_3$ \cite{Arriaga-Osante:2023wnu}, $A_4\times\mathbb{Z}_2$ \cite{Liu:2021gwa}, $2O$ \cite{Ding:2023ydy},  $GL(2,3)$ \cite{Liu:2021gwa}, $\Delta(384)$ \cite{CentellesChulia:2026bkr,Liu:2026ati}. The $\Gamma_N$ are quotients of $\Gamma$ by its principal congruence subgroups $\Gamma(N)$ (for the first values $n$ they are isomorphic to permutation groups \cite{deAdelhartToorop:2011re}) while the $\Gamma_N'$ are their double covering \cite{Ding:2019zxk}. The non-trivial interplay between the modular group and more traditional flavor symmetries acting on the matter fields has been investigated in the context of Froggatt-Nielsen symmetries \cite{Binetruy:1995nt,Dudas:1995eq,Dudas:1996fe,Leontaris:1997vw} and of eclectic symmetries \cite{Baur:2019kwi,Nilles:2020nnc,Nilles:2020kgo,Chen:2021prl,Baur:2022hma,Li:2024pff,Li:2023dvm,Feruglio:2026xxo}. Non-supersymmetric $\Gamma$-invariant field theories have also been considered, see \cite{Qu:2024rns,Ding:2024inn,Qu:2025ddz} in which holomorphy is replaced by a more general harmonicity condition.

The construction of a modular-invariant Lagrangian then requires that one can catalog all arrangements of the fields compatible with the modular symmetry, in particular one ought to determine which functions of $\tau$ can be contracted with matter fields into singlets of $G_d$ of vanishing weight. Since we assumed supersymmetry, these functions are necessarily holomorphic and, under some assumptions on their regularity and boundedness, they are known as Vector-Valued Modular Forms (VVMFs) \cite{Liu:2021gwa}. Similarly to matter fields, they are characterized by a weight and an irreducible representation of $G_d$, as we review in Section~\ref{sec:VVMF}. Therefore, in order to work out the precise predictions of a given model, one needs to evaluate all relevant VVMFs at the relevant vacuum expectation value of the modulus, and thus to have access to their explicit expressions. In the physics literature, closed-form expressions or appropriate Taylor expansions of the VVMFs have often been built case by case, when they were needed for the construction of a new field-theoretic model. For this reason, the special finite modular groups $\Gamma_N'$ have naturally been first used to build models, as it is known how to easily get access to their VVMFs: the lowest-weight ones are built via tractable techniques (see e.g. Ref.~\cite{Feruglio:2017spp}), then their tensor products generate all other VVMFs. 

Although this process has now been repeated for an extensive list of finite modular groups $G_d$ and group representations, it must be performed again for the many interesting cases that remain to be studied. Furthermore, one may wish to systematically scan over vast classes of groups and/or models, in order e.g. to look for universal patterns or to pinpoint remarkable behaviors. For instance, by identifying a group-theoretic pattern common to all groups studied in the past literature, Ref.~\cite{Feruglio:2022koo,Feruglio:2023mii} suggested that only some remarkable values of the modulus can reproduce the observables of neutrino physics. Validating such a prediction in actual models requires the knowledge of many VVMFs of many groups, in particular since non-linear relations between VVMFs sometimes generate cancellations. This generally holds for models relying on hierarchies that naturally emerge close to fixed points of the modular groups \cite{King:2020qaj,Okada:2020ukr,Feruglio:2021dte,Novichkov:2021evw}. Studying as many cases as possible requires that one systematizes and automatizes the construction of VVMFs, which is what we deal with in this paper. We start from two observations:
\begin{itemize}[noitemsep]
    \item in all known cases, there exists a small set of VVMFs of low weights that, once they have been built, can be algebraically combined to form all others through the tensor products of their group representations,
    \item explicit formulae are known for VVMFs of low weights and small representations \cite{Liu:2021gwa}.
\end{itemize}
This suggests an algorithm which first constructs the relevant low-weight VVMFs, using the rules presented in Ref.~\cite{Liu:2021gwa} and reviewed in Section~\ref{sec:VVMF}, then multiplies them in order to generate all others. An obvious question then is: how to identify from the start the relevant low-weight VVMFs? To address this problem, we identify a necessary criterion, purely based on representation theory and discussed in Section~\ref{sec:RepTheoryCriterion}, which actually appears to be sufficient for the groups studied in the past literature and several other which we consider below. Thanks to the Burnside-Brauer theorem, we phrase this criterion in Section~\ref{sec:FaithfulnessAndBB} in terms of the faithfulness of a representation, which is conveniently captured by its character. The resulting algorithm has been implemented in a Python-Mathematica code, available at~\cite{GitHub_vsaulquin}, documented in Section~\ref{sec:CodeStructure}, and applied in Section~\ref{sec:Benchmark model} to the construction of a model of leptons based on the new finite modular group $A_4\times\mathbb{Z}_4$. 

Since our criterion has been established based on observations and scans, but has not yet been proven to be sufficient in all possible cases, we carefully list all potential caveats that might lead to missing VVMFs in Sections~\ref{sec:RepTheoryCriterion} and ~\ref{sec:CodeStructure}. Importantly, we have included all associated checks and warnings in our code, so that potential failures of our criterion for an exotic group not yet included in our database would automatically be notified to the code user. Were it to happen, the expressions produced by the code always remain correct and can be used for model building, yielding a model that is genuinely modular-invariant, albeit tuned in an unjustified way.

\section{Modular-invariant field theories}\label{sec:modInvFT}

We begin with a review of modular-invariant supersymmetric\footnote{We focus on global supersymmetry. The changes brought by supergravity -- such as modular symmetry becoming an $R$-symmetry -- can be found e.g. in \cite{Liu:2026ati,Feruglio:2023uof}.} field theories with a single modulus \cite{Ferrara:1989bc,Ferrara:1989qb,Feruglio:2017spp}. They are invariant under the modular group $\Gamma\equiv SL(2,\mathbb{Z})$ which is the special linear group of integer matrices:
$$SL(2,\mathbb{Z})=\left\{
\left.
\begin{pmatrix}
a&b\\
c&d
\end{pmatrix}\right|a,b,c,d\in\mathbb{Z}\text{ and }ad-bc=1
\right\}$$
It has two generators $S$ and $T$ satisfying the following presentation:
$$SL(2,\mathbb{Z})=\left<S,T|S^4=(ST)^3=S^2TS^{-2}T^{-1}=1\right>$$
where 
$$S=\begin{pmatrix}
    &0&1\\
    &-1&0
\end{pmatrix}
,\quad T=
\begin{pmatrix}
    1&1\\
    0&1
\end{pmatrix}$$
Note that according to the presentation of the group, the element $S^2$ commutes with every other element. By Schur's lemma, for any irreducible representation $\rho$ of $\Gamma$, $\rho(S^2)$ is proportional to the identity. Since $S^4=1$, there are only two possibilities $\rho(S^2)=\pm I_d$. Representations with $\rho(S^2)=I_d$ are called even while the other are said to be odd. 

Modular-invariant theories often introduce at least one modulus $\tau$ on which $\Gamma$ acts non-linearly, whose background value thus spontaneously breaks the modular group. The modulus takes values on the Poincaré half-plane\footnote{One may compactify $\cal H$ by appending to it the point $\tau=i\infty$, where modular forms are required to remain holomorphic (see below).} $\mathcal{H}=\{\tau\in\mathbb{C},\Im(\tau)>0\}$, on which $\Gamma$ acts as follows:
$$\tau\xrightarrow{\gamma}\frac{a\tau+b}{c\tau+d},\quad\text{where }\gamma=
\begin{pmatrix}
    a&b\\
    c&d
\end{pmatrix}
\in\Gamma$$
Note that $S^2\tau=\tau$, hence $\tau$ effectively feels $PSL(2,\mathbb{Z})=SL(2,\mathbb{Z})/\{I_2,-I_2\}$. However, $S^2$ can have a non trivial action on other fields of the theory, as will be emphasized later. Up to the action of $\Gamma$, $\tau$ can always be taken within the \textit{fundamental domain}, i.e. the subset of $\cal H$ such that $-\frac{1}{2}<\text{Re}(\tau)\leq\frac{1}{2}$ and $|z|>1$ or $|z|=1$ and $0\leq\text{Re}(\tau)\leq\frac{1}{2}$.

In addition to the modulus $\tau$, the theory may contain a set of chiral supermultiplets $\Phi_I$, on which the modular group is assumed to act according to:
$$\Phi_I\xrightarrow{\gamma}(c\tau+d)^{-k_I}\rho_I(\gamma)\Phi_I,\quad\gamma=
\begin{pmatrix}
    a&b\\
    c&d
\end{pmatrix}\in SL(2,\mathbb{Z})$$
where $-k_I$ is the weight of the superfield $\Phi_I$ and $\rho_I$ is usually taken to be a unitary representation of a finite modular group. 

A finite modular group $G_d$ is a finite quotient group of $\Gamma$, namely there exists a normal subgroup $H$ of finite index of $\Gamma$ such that $G_d=\Gamma/H$. It accepts a presentation of the form:
$$G_d=\left<S,T|S^4=(ST)^3=S^2TS^{-2}T^{-1}=1+\text{other relations}\right>$$
For instance the usual finite modular groups denoted $\Gamma_N'$ are the quotient groups of $\Gamma$ with the normal subgroups known as the principal congruence subgroups of level $N$:
$$\Gamma(N)=\left\{
\begin{pmatrix}
a&b\\
c&d
\end{pmatrix}\in\Gamma,a\equiv d\equiv 1\pmod{N},b\equiv c\equiv 0\pmod{N}
\right\}$$
They have for $N \leq5$ the following presentation:
$$\Gamma_N'=\left<S,T|S^4=(ST)^3=S^2TS^{-2}T^{-1}=T^N=1\right>$$
Irreducible representations of $G_d$ are irreducible representations of $\Gamma$, thus a $G_d$-invariant theory is automatically $\Gamma$-invariant (but not the other way around). The non-abelian finite modular groups of lowest orders are listed in Table~\ref{Finite modular groups}.
\begin{table}
    \centering
    \begin{tabular}{|c|c|c|}
    \hline
     Cardinal & Additional relations & Finite modular group \\
     \hline
     $6$ & $T^2$ & $S_3$\\
    \hline
    $12$ & $S^2T^2$ & $2D_3$\\
    \hline
    $12$ & $S^2,T^3$ & $A_4$\\
    \hline
    $18$ & $ST^{-2}ST^2$ & $S_3\times\mathbb{Z}_3$\\
    \hline
    $24$ & $T^3$ & $T'$\\
    \hline
    $24$ & $S^2,T^4$ & $S_4$\\
    \hline
    $24$ & $S^2,(ST^{-1}ST)^2$ & $A_4\times\mathbb{Z}_2$\\
    \hline
    $36$ & $S^3T^{-2}ST^2$ & $(\mathbb{Z}_3\rtimes\mathbb{Z}_4)\times\mathbb{Z}_3$\\
    \hline
    $42$ & $T^6,(ST^{-1}S)^2TST^{-1}ST^2$ & $\mathbb{Z}_7\rtimes\mathbb{Z}_6$\\
    \hline
    $48$ & $S^2T^4$ & $2O$\\
    \hline
    $48$ & $T^8,ST^4ST^{-4}$ & $GL(2,3)$\\
    \hline
    $48$ & $T^4$ & $S_4'$\\
    \hline
    $48$ & $(ST^{-1}ST)^2$ & $A_4\times\mathbb{Z}_4$\\
    \hline
    $48$ & $S^2(ST^{-1}ST)^2$ & $T'\times\mathbb{Z}_2$\\
    \hline
    $48$ & $T^{12},ST^3ST^{-3}$ & $((\mathbb{Z}_4\times\mathbb{Z}_2)\rtimes\mathbb{Z}_2\rtimes\mathbb{Z}_3$\\
    \hline
    $54$ & $T^6,(ST^{-1}ST)^3$ & $(\mathbb{Z}_3\times\mathbb{Z}_3)\rtimes\mathbb{Z}_6$\\
    \hline
    $60$ & $S^2,T^5$ & $A_5$\\
    \hline
    $72$ & $T^{12},ST^4ST^{-4}$ & $S_4\times\mathbb{Z}_3$\\
    \hline
    $72$ & $S^2,T^6,(ST^{-1}STST^{-1}S)^2T2$ & $A_4\times S_3$\\
    \hline
    $78$ & $T^6,(ST^{-1}S)^2(TST¨{-1}S)^2T^2$ & $\mathbb{Z}_{13}\rtimes\mathbb{Z}_6$\\
    \hline
    $96$ & $T^8,(ST^{-1}ST)^3$ & $\Delta(96)$\\
    \hline
    $96$ & $T^8,STST^{-2}STS^{-1}T^{-2}$ & $T^{}\rtimes\mathbb{Z}_4$\\
    \hline
    $96$ & $T^{12},TST^{-3}S^{-1}T^2$ & $T'\times\mathbb{Z}_4$\\
    \hline
    $96$ & $S^2,T^6,(ST^{-1}ST)^4$ & $((\mathbb{Z}_4\times\mathbb{Z}_4)\rtimes\mathbb{Z}_2)\rtimes\mathbb{Z}_3$\\
    \hline
    $120$ & $T^5$ & $A_5'$\\
    \hline
    $384$ & $T^{16},(ST^{-1}ST)^3$ & $\Delta(384)$\\
    \hline
    \end{tabular}
    \caption{Non-abelian finite quotient groups of $SL(2,\mathbb{Z})$ of cardinal less than 72 \cite{Liu:2021gwa} extended with every quotient group of $PSL(2,\mathbb{Z})$ \cite{ConderDobcsanyi2006} some of $SL(2,\mathbb{Z})$ of cardinal up to 100. $A_5'$ and $\Delta(384)$ have been added to the list since they have been explored in the literature.}
    \label{Finite modular groups}
\end{table}

 The Lagrangian is the one of a $\mathcal{N}=1$ global supersymmetric theory:
$$\mathcal{L}=\int \text{d}^2\theta\text{d}^2\bar{\theta}\mathcal{K}(\tau,\bar{\tau},\Phi_I,\bar{\Phi}_I)+\left(\int\text{d}^2\theta W(\tau,\Phi_I)+\text{h.c.}\right)$$
 The Kähler potential is chosen to be minimal
 \footnote{This assumption is standard in the literature, although it is known that non-minimal Kähler potentials can have a strong impact on the parameter counting of the theory and on its predictive nature \cite{Chen:2019ewa}. For models of flavor physics which rely on natural exponential hierarchies at the vicinity of the points $\tau=i,\omega$ or $i\infty$, Kähler corrections reduce accidental suppressions beyond the predicted hierarchies. Ecletic flavor symmetries \cite{Baur:2019kwi,Nilles:2020nnc,Nilles:2020kgo,Chen:2021prl,Baur:2022hma,Li:2024pff,Li:2023dvm,Feruglio:2026xxo} may control the Kähler corrections. Our focus here is the systematic exploration of VVMFs of any finite modular group, so we do not investigate the topic of Kähler corrections further.
 }:
\begin{equation}\label{eq:Kahler}
\mathcal{K}(\tau,\bar{\tau},\Phi_I,\bar{\Phi}_I)=-h\log(-i\tau+i\bar{\tau})+\sum_I(-i\tau+i\bar{\tau})^{-k_I}|\Phi_I|^2    
\end{equation}
with $h>0$ a positive constant. This Kähler potential is invariant under modular transformation up to a Kähler transformation, $\mathcal{K}\xrightarrow[]{\gamma}\mathcal{K}+F(\tau,\Phi_I)+\bar F(\bar\tau,\bar\Phi_I)$ for the holomorphic function $F=h\log(c\tau+d)$. The superpotential can be expressed in power series of the superfields:
$$W(\tau,\Phi_I)=\sum_nY_{I_1,\dots,I_n}(\tau)\Phi_{I_1}\dots\Phi_{I_n}$$
where, importantly, the Yukawa couplings $Y_{I_1,\dots,I_n}$ can be functions of $\tau$ instead of simple coupling constants \cite{Dixon:1986qv,Erler:1991nr,Dixon:1989fj}. For the theory to be modular invariant, the superpotential must itself be modular invariant (in addition to being gauge invariant when the superfields transform under a gauge group, and/or invariant under other more traditional global symmetries). This requires that the Yukawa couplings transform as:
$$\forall \gamma\in\Gamma, \ Y_{I_1,\dots,I_n}(\gamma\tau)=(c\tau+d)^{k_Y}\rho_Y(\gamma)Y_{I_1,\dots,I_n}(\tau)$$
where $k_Y$ is the weight and $\rho_Y$ is a representation of the finite modular group $G_d$, and that
$$k_Y=k_{I_1}+\cdots+k_{I_n},\quad\rho_Y\otimes\rho_{I_1}\otimes\dots\otimes\rho_{I_n}\supset\mathbf{1}$$
where \textbf{1} is the trivial representation.
Beyond these conditions which ensure that the functions $Y(\tau)$ complete the fields into a modular-invariant superpotential term, one usually further requires that they are holomorphic everywhere on the compactification of the Poincaré half-plane. This makes them \textit{Vector-Valued Modular Forms}, which we now discuss.

\section{Vector-Valued Modular Forms}
\label{sec:VVMF}

We now review the pieces of mathematics of VVMFs that are relevant for our purposes. We mostly follow \cite{Liu:2021gwa}, which applied this formalism to flavor physics, thereby extending flavor model building to unexplored finite modular groups beyond the well-known $\Gamma_N$.

A VVMF is an holomorphic function of the modulus $\tau$ whose image is a vector of complex numbers:
\begin{equation}
\label{eq:x}
Y:\left |
\begin{aligned}
\mathcal{H} &\longrightarrow \mathbb{C}^d 
\\
\tau &\longmapsto (Y_1(\tau),\dots,Y_d(\tau))^T \
\end{aligned}
\right.
\end{equation}
which is bounded as $\text{Im}(\tau)\to \infty$ and which satisfies the relation:
\begin{equation}
\label{eq:VVMF transfo}
\forall\gamma\in SL(2,\mathbb{Z}),\quad Y(\gamma\tau)=(c\tau+d)^k\rho(\gamma)Y(\tau)
\end{equation}
where $k$ is a non-negative integer called the weight of the VVMF and $\rho$ is a representation of $SL(2,\mathbb{Z})$. That the weight should be non-negative follows from the requirement that the form is well-behaved at $\tau\to i\infty$. That it is integer is required by the action of $S^2$: taking $\gamma=S^2$ in the former expression leads to $Y(\tau)=(-1)^k\rho(S^2)Y(\tau)$, so $(-1)^k\rho(S^2)=I_d$. $k$ thus takes only integral values and even/odd weight VVMFs can only transform under even/odd representations, respectively.

For a representation $\rho$ of a group $G$, $\text{Ker}(\rho)$ is a normal subgroup of $G$ and $\rho$ defines a representation of $SL(2,\mathbb{Z})/\text{Ker}(\rho)$. Given our assumptions of Section~\ref{sec:modInvFT} regarding how fields transform, we focus on cases where  
$\text{Ker}(\rho)$ is of finite index, such that $G_d=SL(2,\mathbb{Z})/\text{Ker}(\rho)$ is a finite modular group. 
Finally, since for a group morphism $\text{Im}(\rho)\cong G/\text{Ker}(\rho)$, the assumption that $\text{Ker}(\rho)$ is of finite index translates into the fact that $\text{Im}(\rho)$ is finite. This assumption together with Maschke theorem ensure that $\rho$ can be restricted to a unitary irreducible representation, as we will now assume.

The transformation \eqref{eq:VVMF transfo} of a VVMF retains the same shape after a change of orthonormal basis, but the entries of each $\rho(\gamma)$ are shuffled. A convenient basis for expressing the VVMF is that where $\rho(T)$ is diagonal. Since $\text{Im}(\rho)$ is finite, there exists $N$ such that $\rho(T)^N=1$. Then one can write
\begin{equation}
\label{eq:diagT}
\rho(T)=\exp(2i\pi L)    
\end{equation}
where $L=(r_1,\dots,r_d)$, with $r_i \in \frac 1 N  \mathbb Z$ and the conventional choice that $0\leq  r_i<1$. When $\gamma=T$ in \eqref{eq:VVMF transfo}, the VVMFs must satisfy the following relation:
$$Y(\tau+1)=\rho(T)Y(T)$$
In the basis where $\rho(T)$ is diagonal, the $i$-th component of the VVMF thus satisfies $Y_i(\tau+1)=\exp(2i\pi r_i)Y(\tau)$. This shows that $Y_i$ is a quasiperiodic function of $\tau$, hence it is amenable to a Fourier decomposition, known as a $q$-expansion, where $q=e^{2i\pi \tau}$:
$$Y_i(\tau)=q^{r_i}\sum_{n\geq0}a_i(n)q^n$$
Note that $n\geq0$ since $Y(\tau)$ is bounded as $\text{Im}(\tau)\rightarrow \infty$. The goal of this paper is to algorithmically construct $q-$expansions of VVMFs.

\subsection{Module Structure}

Let $\mathcal{M}_k(\rho)$ be the set of VVMFs of weight $k$ and transforming under the representation $\rho$. It clearly is a vector space, moreover one can show that for all $k>1$ \cite{candelori2016vectorvaluedmodularforms}:
\begin{equation}
\label{eq:dimMKrho}
\dim(\mathcal{M}_k(\rho))=\frac{(5+k)\dim(\rho)}{12}+\frac{i^k\text{Tr}(\rho(S^3))}{4}+\frac{(-\omega^2)^k\text{Tr}(\rho(S^3T))}{3(1-\omega)}+\frac{\omega^k\text{Tr}(\rho((ST)^2))}{3(1-\omega^2)}-\text{Tr}(L)    
\end{equation}
Similarly $\mathcal{M}(\rho)=\oplus_{k=1}^{\infty}\mathcal{M}_k(\rho)$ denotes the whole space of modular forms transforming under $\rho$. With these notations $\mathcal{M}(\bf 1)$ is the set of the usual modular forms i.e. the holomorphic functions on $\mathcal{H}$ bounded at imaginary infinity satisfying $Y(\gamma\tau)=(c\tau+d)^kY(\tau)$. A classical result \cite{cohen2017modular} is that $\mathcal{M}(\mathbf{1})=\mathbb{C}[E_4,E_6]$, meaning that every modular form is a weight-homogeneous polynomial of the two Eisenstein series $E_4$ and $E_6$ defined by
\begin{equation}
\label{eq:DefEinsensteinSeries}
E_k(\tau)=1-\frac{2k}{B_k}\sum_{n=1}^\infty\sigma_{k-1}(n)q^n \ ,    
\end{equation}
where $q=e^{2i\pi\tau}$, $B_k$ is the $k$-th Bernoulli number and $\sigma_k(n)=\sum_{d|n}d^k$.
(Note that this formula defines any Eisenstein series for $k$ a non-negative even integer.) $E_4$ and $E_6$ have weight 4 and 6, respectively.
The free-module theorem \cite{Marks_2010} states that $\mathcal{M}(\rho)$ is a free module over $\mathcal{M}(\bf1)$ of rank $d=\text{dim}(\rho)$, namely there exists $d$ VVMFs $F_1,\dots,F_d$ in $\mathcal{M}(\rho)$ such that every VVMF $Y$ in $\mathcal{M}(\rho)$ can be written:
\begin{equation}
\label{eq:moduleStructure}
Y(\tau)=\sum_{i=1}^dP_i(\tau)F_i(\tau)
\end{equation}
where every $P_i$ is an element of $\mathbb{C}[E_4,E_6]$ made out of monomials of identical weights such that the sum of the weights of $F_i$ and $P_i$ is equal to the weight of $Y$.

\subsection{Basis of the module}\label{sec:BasisVVMF}
Cataloging the VVMFs transforming under a given representation $\rho$ thus reduces to determining the basis $F_1,\dots,F_d$ of $\mathcal{M}(\rho)$. Fortunately, there exist explicit formulae for the lowest dimensions $d$.

Let us first introduce the modular derivative $D_k$ \cite{bruinier2008one}: for a given VVMF $Y$ of weight $k$ and representation $\rho$, its modular derivative is defined as:
\begin{equation}
\label{eq:modularDerivative}
D_kY\equiv\frac{1}{2i\pi\tau}\frac{\dd Y}{\dd\tau}-\frac{kE_2}{12}Y \ ,    
\end{equation}
where $E_2$ is defined in \eqref{eq:DefEinsensteinSeries}. Unlike $E_{4,6}$, $E_2$ is not a modular form, it transforms as $E_2(\gamma\tau)=(c\tau+d)^2E_2(\tau)-\frac{6i}{\pi}c(c\tau+d)$. The non-linear shift of $E_2$ under a modular transformation cancels that of $\frac{\dd Y}{\dd \tau}$ such that $D_kY\in\mathcal{M}_{k+2}(\rho)$ if $Y\in\mathcal{M}_k(\rho)$. Moreover, the modular derivative satisfies the Leibniz rule i.e. if $Y_1,Y_2$ are modular forms of weights $k_1,k_2$ respectively then 
$$D_{k_1+k_2}(Y_1\otimes Y_2)=D_{k_1}(Y_1)\otimes Y_2+Y_1\otimes D_{k_2}(Y_2)$$
We will apply the modular derivative several times below, hence we introduce the shorthand:
$$D_k^n=D_{k+2(n-1)}\circ\cdots\circ D_k$$

For representations of dimension $d$ up to 3, a basis of ${\cal M}_k(\rho)$ can be built out of a minimal VVMF of weight $k_0$, denoted $F_{k_0}$, and as its $d-1$ first modular derivatives.
Then, according to \eqref{eq:moduleStructure} a modular form of weight $k$ writes
  $F(\tau)= \sum _{n=0}^{d-1} P_n(E_4,E_6) D_{k_0}^n F_{k_0}(\tau)$ where
  $P_n(E_4,E_6)$ are polynomials in $E_4,E_6$ of weight $k-k_0-2n$.
In these cases, it is said that $\mathcal{M}(\rho)$ is cyclic. The minimal weight $k_0$ can be explicitly computed from the row $L=(r_1,\dots,r_d)$ defined in Eq.~\eqref{eq:diagT}, while the minimal VVMF can be expressed in terms of $L$, the Dedekind eta function $\eta$, a local parameter $K(\tau)$ and the generalized hypergeometric series ${}_{n}F_{n-1}$ (see Appendix \ref{sec:Useful functions} for definitions). In the basis of Eq.~\eqref{eq:diagT} where $\rho(T)$ is diagonal, one finds \cite{zbMATH06664578,franc2015structuremodulesvectorvalued}:
\begin{itemize}
\item For $\dim(\rho)=1$, the basis is $F_{k_0}=\eta^{2k_0}$ with $k_0=12\text{Tr}(L)$.

\item For $\dim(\rho)=2$, the basis is $(F_{k_0},D_{k_0}F_{k_0})$ with $k_0=6\text{tr}(L)-1$.  Here $F_{k_0}$ has the following expression:
\begin{equation}
\label{Dim2VVMFformula}
F_{k_0}=
\left(
\begin{matrix}
\eta^{2k_0}K^{\frac{6(r_1-r_2)+1}{12}}{}_{2}F_1\left(\frac{6(r_1-r_2)+1}{12},\frac{6(r_1-r_2)+5}{12};r_1-r_2+1;K\right)\\
C_2\eta^{2k_0}K^{\frac{6(r_2-r_1)+1}{12}}{}_{2}F_1\left(\frac{6(r_2-r_1)+1}{12},\frac{6(r_2-r_1)+5}{12};r_2-r_1+1;K\right)
\end{matrix}
\right)
\end{equation}

\item For $\dim(\rho)=3$, the basis is $(F_{k_0},D_{k_0}F_{k_0},D_{k_0}^2F_{k_0})$ with $k_0=4\text{Tr(L)}-2$.Here $F_{k_0}$ has the following expression:
\begin{equation}
\label{Dim3VVMFformula}
F_{k_0}=
\left(
\begin{matrix}
\eta^{2k_0}K^{\frac{a_1+1}{6}}{}_{3}F_2\left(\frac{a_1+1}{6},\frac{a_1+3}{6},\frac{a_1+5}{6};r_1-r_2+1,r_1-r_3+1;K\right)\\
C_3\eta^{2k_0}K^{\frac{a_2+1}{6}}{}_{3}F_2\left(\frac{a_2+1}{6},\frac{a_2+3}{6},\frac{a_2+5}{6};r_2-r_3+1,r_2-r_1+1;K\right)\\
C_3'\eta^{2k_0}K^{\frac{a_3+1}{6}}{}_{3}F_2\left(\frac{a_3+1}{6},\frac{a_3+3}{6},\frac{a_3+5}{6};r_3-r_1+1,r_3-r_2+1;K\right)
\end{matrix}
\right)
\end{equation}
with $a_1=4r_1-2r_2-2r_3$, $a_2=4r_2-2r_1-2r_3$ and $a_3=4r_3-2r_1-2r_2$.
\end{itemize}
The constants $C_2,C_3,C_3'$ are fixed by the requirement that $F_{k_0}$ transforms appropriately under the modular action of $\gamma=S$. An important remark is that these expressions in terms of the generalized hypergeometric series are valid for $\tau$ in the fundamental domain. However, this is enough to compute the coefficients of the $q$-expansions, which then hold on the whole Poincaré half-plane.

The case $\dim(\rho)=4$ is more subtle since there are two distinct cases \cite{franc2018constructionsvectorvaluedmodularforms}. Only when $\rho(S^2)=-(-1)^{3\text{Tr}(L)}\text{Id}_4$ is the representation cyclic. In this case, there is a unique minimal VVMF $F_{k_0}$ of weight $k_0=3\text{Tr}(L)-3$ and the basis is $(F_{k_0},D_{k_0}F_{k_0},D_{k_0}^2F_{k_0},D_{k_0}^3F_{k_0})$. In the non-cyclic case, there are two minimal VVMFs $F_{k_0},G_{k_0+2}$ with $k_0=3\text{Tr}(L)-2$ and the basis is $(F_{k_0},D_{k_0}F_{k_0},G_{k_0+2},D_{k_0}^2F_{k_0})$. In both cases, there is no explicit expression with generalized hypergeometric series for the minimal VVMFs but it is possible to recursively get the coefficients of the $q$-expansion for the lowest weight VVMF $F_{k_0}$ (see Appendix \ref{sec:Recursive method} for more details).

When $\dim(\rho)=5$, there are five options for $k_0$ \cite{Liu:2021gwa}. When $\dim(\rho)\geq 6$, there are partial results \cite{franc2015structuremodulesvectorvalued}, but we are not aware of expressions explicit enough for our purposes.

\section{The minimal set of VVMFs}

Thanks to the formalism reviewed in Section~\ref{sec:VVMF} one can build the VVMFs of every finite modular group. Indeed, one browses each irreducible representation $\rho$ of the group, builds the lowest-weight basis of the module ${\cal M}(\rho)$ through the explicit or recursive expressions presented above, then generates the higher-weight VVMFs by multiplying the basis by polynomials of the Einsenstein functions $E_4$ and $E_6$. 

This method yields all VVMFs representation by representation. However, it has some practical drawbacks. First, for field theory model building, it is preferable to know all VVMFs weight by weight, since the weights of relevant VVMFs are fixed by the model's interactions, while all representations leading to flavor-invariant interactions shall be included. Second, if the finite modular group has high-dimensional (in practice, $d\geq 5$) irreducible representations, the expressions of the lowest-weight VVMFs from which the whole set is generated are not known.

What is instead commonly done to compute the VVMFs of a finite modular group is to start from a minimal set of low-weight VVMFs and to perform successive tensor products to generate the higher-weight ones. This method has the advantage of needing only a few minimal VVMFs, instead of one by representation, and it can in principle reach every representation that arises from the tensor products, whatever its dimension. It however relies on the ability to identify the relevant set of low-weight VVMFs, and to subsequently build them, given that expressions are known for usual finite modular groups but can be difficult to find for more exotic ones. 

The purpose of this section is to systematically combine these two approaches in order to build every VVMF for any finite modular group: the minimal set of VVMFs is computed\footnote{As said above, this fails when the minimal VVMFs contain high-dimensional representations, in which case one shall rely on other, e.g. recursive, techniques.} via the formalism of Section~\ref{sec:BasisVVMF}, and the other ones are generated from tensor products. However, the first question that needs to be adressed is: how does one determine the minimal set of VVMFs ? We start by some examples to illustrate the subtleties which may arise. 

\subsection{Exhibiting a representation-theoretical criterion}
\label{sec:RepTheoryCriterion}

As a first example we focus on the modular group $\Gamma_3=A_4$. The lowest-weight VVMFs of $A_4$ were studied and built in Ref.~\cite{Feruglio:2017spp}. They have weight 2 and there are three independent ones, which can be organized such that they transform as the irreducible triplet representation of $A_4$. Then, a combinatorial argument shows that tensor products  of these three VVMFs are enough to get all other VVMFs of $A_4$, hence the minimal set for $A_4$ simply corresponds to the lowest-weight VVMFs arranged in a single irreducible representation. This observation still holds for $T'$, the double covering of $A_4$ \cite{Liu:2019khw}. The lowest-weight VVMFs of $T'$ have weight 1 and transform as an irreducible doublet. Here again they generate all VVMFs of the group from tensor products. 

In these two examples, the lowest-weight VVMFs appear in a single irreducible representation, but that is not always so. For instance, the lowest weight for the finite modular groups $GL(2,3)$ 
is 2, and there are seven associated weight-2 VVMFs, organized in two different irreducible doublets and one irreducible triplet. Yet, by forming every possible tensor product between these three multiplets, one generates every VVMF of $GL(2,3)$. Up to this point, the minimal set of VVMFs seems to be that of the lowest-weight ones, whichever set of multiplets they are organized into. This observation extends to all $\Gamma_N^{(')}$ and to some exotic finite modular groups (such as $GL(2,3)$).

However, for other finite modular groups, the identification of their minimal set with that of the lowest-weight VVMFs may fail, 
which we now exemplify. First, let us consider the finite modular group $2O$ \cite{Ding:2023ydy}. Its lowest-weight VVMFs have weight 2, whose tensor products can only produce even-weight VVMFs. However, $2O$ admits odd-weight VVMFs, for example of weight 3. In this case, the minimal set must be extended to the next-to-lowest weight VVMFs, which then suffices since one can check that tensor products of VVMFs of weight 2 and 3 generate all other ones. Thus, some groups' minimal set needs more than the lowest-weight VVMFs, for instance when some of the allowed weights\footnote{The allowed weights can be extracted from \eqref{eq:dimMKrho}, as those such that $\text{dim}({\cal M}_k(\rho))>0$ for at least one $\rho$, i.e. such that $\sum_\rho \text{dim}({\cal M}_k(\rho)) > 0$.} are not multiples of the lowest one, as it happens for $2O$.
Another illuminating example is given by the group $A_4\times\mathbb{Z}_4$. It has a weight-1 VVMF, whose weight divides every possible weight, but it transforms as a non-trivial singlet representation whose tensor products can never transform as the triplet representations of $A_4\times\mathbb{Z}_4$. Generally, one shall extend the minimal set beyond the lowest-weight VVMFs if the set of representations that they span does not generate all of the group's representations from tensor products. Actually, even for $2O$ one can phrase what happens in terms of representations instead of weights: the representation of the lowest-weight VVMFs satisfies $\rho(S^2)=I_d$ (as for any even-weight VVMF), hence their tensor products cannot be such that $\rho(S^2)=-I_d$. The minimal sets for these last two examples can thus be analyzed through the unique lens of whether they contain enough representations to generate every other representation from tensor products. 

This will constitute the backbone of our algorithm to construct the minimal set of VVMFs: we determine the representations under which the lowest-weight VVMFs transform, we stop if these representations are sufficient to generate every other irreducible representation from tensor products, otherwise we determine the representations of the next-to-lowest weight, and we proceed recursively until we manage to form all representations. This demands a criterion to figure out whether a set of representations can generate every other irreducible representation of the group, which is exactly the purpose of the Burnside-Brauer theorem which we discuss in the next section. We stress that, although for the groups listed in Table~\ref{Finite modular groups} the recursive procedure produces a minimal set containing at most the two lowest weights, it can handle a group whose minimal set involves additional weights.

However, note that our criterion is necessary but not sufficient. There could for instance exist a finite modular group $G_d$ whose lowest-weight VVMFs have weight $k$ and transform under a (potentially reducible) representation $\rho$ that generates all other irreducible representations of $G_d$, but such that some VVMF exists in an irreducible representation $\rho'$  with a weight that is strictly less than the number of products of $\rho$ needed to form $\rho'$.
For instance, $\rho'$ may arise after three tensor products of $\rho$, leading to a VVMF of weight $3k$, while there may exist a VVMF transforming under $\rho'$ of weight $k'<3k$.
Then, the minimal set for this group shall contain VVMFs of weights $\tilde k$, with $k<\tilde k\leq k'$. This scenario, although possible, does not arise for the groups that are listed in Table~\ref{Finite modular groups}. Similarly, the minimal set may generate every representation at every allowed weight from tensors products, but it may fail to produce enough independent VVMFs. As said above, associated to a representation $\rho$ and a weight $k$ is a vector space $\mathcal{M}_k(\rho)$ of VVMFs. Tensor products of the minimal set may generate strictly less than $\dim(\mathcal{M}_k(\rho))$ (given by \eqref{eq:dimMKrho}) independent VVMFs of weight $k$ in the representation $\rho$. Note that, due to the module structure of \eqref{eq:moduleStructure}, this case can only happen for finitely-many weights, so a notion of minimal set always exists. We have checked that, for every group for which we have included the Clebsch-Gordan coefficients in our code, it does not happen and that the minimal set effectively generates every VVMF from tensor products. There may exist a mathematical theorem stating that the scenarios described in this paragraph are not realized by finite modular groups, but we are not aware of it.  In any case, as we explain below, our computer algorithm includes sanity checks that unambiguously tell us through \eqref{eq:dimMKrho} whether the construction led to a sufficient set of VVMFs, and warns the user otherwise. As said above, we have not witnessed any warning during our numerous checks. 

\subsection{Faithful representations and the Burnside-Brauer theorem}
\label{sec:FaithfulnessAndBB}
The Burnside-Brauer theorem \cite{burnside1955theory,brauer1964note,steinberg2014burnsidebrauersteinberg} states that a rep $\rho$ of a finite group $G$ can generate every irreducible representation of $G$ if and only if $\rho$ is faithful, i.e. if $\rho(g)=\text{Id}\Leftrightarrow g=e$. For several finite modular groups, the candidate minimal set of VVMFs contains a family of irreducible representations $\rho_1,\dots,\rho_d$, and we want to know if all possible tensor products of elements of this family can reach every irreducible representation. Noticing that every tensor product among representations of the family is the same as every tensor product of the rep $\rho=\rho_1\oplus\cdots\oplus\rho_d$, the Burnside-Brauer theorem tells us that the family is enough if and only if $\rho$ is faithful.

A classical result about representations of finite groups is that a representation $\rho$ of $G$ is faithful if and only if, for every element of the group different from the identity element, its character $\chi_\rho$ takes a value which is not equal to the dimension of the rep:
$$\rho\text{ is faithful }\Longleftrightarrow \forall g\in G,\chi_\rho(g)\neq\chi_\rho(e)=\dim(\rho)$$
Moreover, the character of the representation $\rho=\rho_1\oplus\dots\oplus\rho_d$ simply is the sum of the characters $\chi_{\rho_i}$. In conclusion, in order to know up to which weight we will systematically construct all VVMFs before restricting to their tensor products, we only need to know the character table of the modular group under consideration. For the groups discussed in the previous section, the genuine minimal sets have been identified. Since they generate all VVMFs from tensor products and that there exists at least one VVMF for every irreducible representation of any group\footnote{From \eqref{eq:dimMKrho}, one sees that $\text{dim}\,\mathcal{M}_k(\rho)\sim \frac{k}{12}\text{dim}\,\rho>0$ when $k\to \infty$, all other terms being bounded once $\rho$ is specified.}, minimal sets ought to combine into a faithful representation. Let us check that our examples indeed fulfill the above characterization by characters.

Let us start with $T'$, whose VVMFs can be obtained from tensor products of its weight-1 VVMF \cite{Liu:2019khw}. As reminded in Appendix~\ref{sec:Group theory dictionnary}, it has 7 conjugacy classes, and its lowest-weight VVMFs transform as a doublet \textbf{2} and have weight 1. The character table of $T'$ tells us that $ \chi_\mathbf{2}=(2,-2,0,\omega^2,-\omega,-\omega^2,\omega)$, where $\omega$ is the third root of unity, which shows that only $[e]$ has a character equal to $2$. Hence, \textbf{2} is faithful and every other irreducible representation can be reached from tensor products of \textbf{2}, as must be the case given the role played by the weight-1 doublet VVMF of $T'$.

Now consider $GL(2,3)$. The minimal VVMFs have weight 2 and transform under the irreducible representations $\mathbf{2}$,$\mathbf{2}'$ and $\mathbf{3}$. $GL(2,3)$ has 8 conjuacy classes and the character table gives: $\chi_\mathbf{2} = (2,2,0,0,2,-1,-1,0),\chi_{\mathbf{2}'}=(2,-2,\sqrt{2}i,-\sqrt{2}i,0,-1,1,0)\text{ and }\chi_{\mathbf{3}}=(3,3,1,1,-1,0,0,-1)$. Then the sum representation has: $\chi_{\mathbf{2}+\mathbf{2}'+\mathbf{3}}=(7,3,1+\sqrt{2}i,1-\sqrt{2}i,1,-2,0,-1)$ and consequently is faithful 
, consistently with the fact that the VVMFs of $GL(2,3)$ can be built as tensor products of the weight-2 VVMF \cite{Liu:2021gwa}. Note that $\mathbf{2}'$ alone is faithful, but the tensor products of the associated VVMF have at least weight 4, so one shall add the other VVMFs of weight 2 in the minimal set. Moreover, the tensor products of the VVMF in $\mathbf{2}'$ with itself do not produce every weight-4 VVMF. This justifies why, when a VVMF at a given weight is needed to form a faithful representation, we include all possible VVMFs at that weight in our minimal set, even if they are associated to representations already covered by VVMFs of lower weights.

Finally, for $2O$ and $A_4\times\mathbb{Z}_4$ we have seen above that the lowest-weight VVMF precisely cannot generate all VVMF as they are associated to unfaithful representations. For $2O$, as already mentioned the lowest weight VVMFs have weight 2 and transform under \textbf{2} and \textbf{3}. $2O$ has 8 conjugacy classes and the character table gives: $\chi_\mathbf{2}=(2,2,0,0,2,-1,-1,0)$ and $\chi_\mathbf{3}=(3,3,1,1,-1,0,0,-1)$ so $\chi_{\mathbf{2}+\mathbf{3}}=(5,5,1,1,1,-1,-1,-1)$. The second entry is the same as the first one thus $\mathbf{2}+\mathbf{3}$ is not faithful
. Adding the weight-3 VVMFs that transform under $\hat{\mathbf{4}}$, whose character is $\chi_{\hat{\mathbf{4}}}=(4,-4,0,0,0,-1,1,0)$, one obtains $\chi_{\mathbf{2}+\mathbf{3}+\hat{\mathbf{4}}}=(9,1,1,1,1,-2,0,1)$, so $\mathbf{2}+\mathbf{3}+\hat{\mathbf{4}}$ is faithful, which can be deduced from the fact that weight-2 and -3 VVMFs generate every other VVMF of $2O$ \cite{Ding:2023ydy}. For completeness, the case of $A_4\times\mathbb{Z}_4$ is treated in Section~\ref{sec:Benchmark model}.

\section{Structure and use of the code}
\label{sec:CodeStructure}
We have now described every step of our algorithmic construction of VVMFs' $q$-expansions for a given finite modular group. We first summarize those steps, before we explain how to practically use the code attached to this paper~\cite{GitHub_vsaulquin}, which applies the systematic method.

\subsection{Finding the minimal set}
As we have extensively discussed, the first step is to identify the minimal set of VVMFs needed to build all the others. Based on the condition of faithfulness required by the Burnside-Brauer theorem, the process to determine this minimal set is:
\begin{itemize}
\item Initialize the set of irreducible representations $\{\rho\}= \emptyset$, then the weight $k=1$. 
\item Determine the representations $\rho_1,\cdots,\rho_d$ for which there exist VVMFs of weight $k$.
\item Update the set of irreducible representations according to $\{\rho\}\to\{\rho\}\cup\{\rho_1,\dots,\rho_d\}$ and build the character of their direct sum: $\chi_{\{\rho\}}\equiv\chi_{\oplus\rho\in\{\rho\}}=\sum_{\rho\in\{\rho\}}\chi_\rho$.
\item If there is a conjugacy class such that $\chi_{\{\rho\}}(g)=\chi_{\{\rho\}}(e)$, increase the weight $k$ by one unit and repeat the procedure from the second step
. If not, stop here.

\end{itemize}
The second step of this process requires that, for a given weight, we know which representations admit VVMFs, which relies on the theory of VVMFs presented above. For a weight $k$, if $\dim(\mathcal{M}_k(\rho))>0$, we know that the representation $\rho$ appears at this weight. If $k>1$, one can directly apply the formula \eqref{eq:dimMKrho}. It does not hold for $k=1$ but, for a given representation, we can compute its minimal weight $k_0$ defined in Section~\ref{sec:BasisVVMF}
 and check whether it is $1$. An important remark is that $k_0$ is only unambiguously known for representations 
of dimension up to $5$,  hence we are currently not able to detect when a high-dimensional representation admits a VVMF of weight 1. In this special case, the algorithm may fail to construct all VVMFs for the group under consideration. However, as said above, the code is designed so as to warn the user if there are missing VVMFs at weights $k>1$, which are straightforward to detect and could suggest a missing VVMF at weight 1. Furthermore, an automatic warning is displayed whenever one uses a group that admits large odd representations, for which a weight-1 VVMF could have been missed. (Note however that a ``fake" minimal set containing weights 2 and 3 would succeed in reproducing all higher-weights tensor products of the missing weight-1 VVMF. This would happen for $\Gamma_5'$ which admits a weight-1 sextuplet VVMF.)

The outcome of this first procedure is the weight up to which we have to build the VVMFs, before we can generate all the other ones from tensor products.

\subsection{Building the VVMFs of the minimal set}

At this point, the minimal set is fully determined, i.e. it is known that it is made of VVMFs of weight $k_1,\dots,k_r$ and that for each weight $k_m$ with $1\leq m\leq r$ there exist $i_m$ VVMFs in the representations $\rho_{k_m}^{i}$ with $1\leq i\leq i_m$. One now ought to explicitly build these VVMFs. 
Seemingly, nothing forbids that a representation appears several times in the minimal set at different weights. If it happens, the VVMF with the lowest of those weights simply is the minimal-weight VVMF for this representation, by construction of the minimal set. 
  Its $q$-expansion is then derived thanks to the expression given in Section~\ref{sec:BasisVVMF} or by recursive methods described in Appendix~\ref{sec:Recursive method}. However, the VVMFs of identical representation but higher weights cannot be obtained from those expressions. In cyclic cases, they are the modular derivatives of the minimal-weight VVMF, while in other cases adapted formulae must be used. Since we are not aware of the existence of such formulae, we only have implemented the procedure for cyclic cases, and required that the code warns the user that some VVMFs are missing when the case is non-cyclic. Note that this discussion is hypothetical since, for the 26 groups in Table~\ref{Finite modular groups}, the minimal set never contains VVMFs that are not the minimal-weight ones for their representation.

\subsection{Tensor products}
Finally, one generates every VVMF of weight $k$, up to a value of interest e.g. in model building, from tensor products of those in the minimal set. We have to form every possible tensor product of overall weight $k$, which produces representations
$$(\rho_{k_1}^1)^{n_{k_1}^1}\otimes\cdots\otimes(\rho_{k_1}^{i_1})^{n_{k_1}^{i_1}}\otimes\cdots\otimes(\rho_{k_r}^1)^{n_{k_r}^1}\otimes\cdots\otimes(\rho_{k_r}^{i_r})^{n_{k_r}^{i_r}}$$
such that
$$\sum_{m=1}^r\sum_{l=1}^{i_m}n_{k_m}^lk_m=k$$
The decomposition of such a tensor product in irreducible representations gives the expected VVMFs of weight $k$. Generally, the decomposition of every possible tensor product produces VVMFs in a number which overshoots $\dim{\cal M}_k(\rho)$ for the appropriate $\rho$, so not all VVMFs thus produced are linearly independent. The next step is then to eliminate all redundant VVMFs in order to only deal with linearly-independent ones, using the fact that VVMFs are linearly-independent as soon as the vectors of coefficients of their $q$-expansions are linearly-independent at some order in $q$. We check at the end of the process that for each representation $\rho$ we have managed to form $\dim\mathcal{M}_k(\rho)$ independent VVMFs, otherwise the user of the code is notified. Finally, we factorize each component of the VVMFs such that its $i$-th entry has the form $C_iq^{r_i} \sum_{n\geq 0}a_i(n)q^n$ where $a_i(n)$ are integers. Since all VVMFs are defined up to an overall multiplicative constant, we normalize them such that every $|C_i|$ is greater or equal than one and at least one $C_i$ is equal to one. See \cite{Petcov:2023fwh} for other choices of normalization.

\subsection{How to use the code}
Our code automatizes the steps described above, i.e. it builds every VVMF of the finite modular group that the user is interested in. More precisely, for a given group and a given weight, it applies the algorithm described in this paper to compute every VVMF of this group at the specified weight. To do so, the code contains a file \texttt{GroupTheoryDictionnary.py}, which is a dictionary listing useful information about the representation theory of every group identified in Table~\ref{Finite modular groups}. The information is gathered in a list \texttt{Group\_Theory}. Let us first describe what it contains, as this will allow us to describe what the algorithm requires and what the user needs to provide to the code to study a group beyond the ones we pre-coded.
\begin{itemize}
\item The first entry of \texttt{Group\_Theory} is a list of lists containing the names of the irreducible representations gathered by dimension. Concretely, the first list names all dimension-1 irreducible representations, the second list all dimension-2 irreducible representations, and so on.
\item The second entry is a list of lists arranged as the previous one, such that the $i$-th element of the $j$-th list is the matrix $\rho(S)$ of the $i$-th irreducible representation of dimension $j$.
\item The third entry is the analog of the second for $\rho(T)$. An important remark is that the theory of VVMFs is developed in the basis where $\rho(T)$ is diagonal, so the code assumes that $\rho(S)$ and $\rho(T)$ are given in this basis.
\item The fourth entry is simply an integer corresponding to the level of the group i.e. the smallest non-zero integer such that $T^N=1$.
\item The fifth entry is a list of integers gathering the sizes of the conjugacy classes of the group.
\item The sixth entry is the character table of the group. It has the same structure as the first entry, namely the $i$-th element of the $j$-th list is the character table of the $i$-th irreducible representation of dimension $j$. The character table is itself a list ordered according to how the conjugacy classes are ordered in the previous entry.
\item The seventh entry just is the \texttt{GAPId} of the group.
\item The eighth entry is the name of the group.
\end{itemize}
If the user wants to use a new group beyond those of Table~\ref{Finite modular groups} which we already tabulated, the expected eight pieces of information can be easily obtained from the software \texttt{GAP} \cite{GAP4102} and copied in a new \texttt{Group\_Theory} object. Examples of the required syntax are provided by the groups already coded. Once the new object \texttt{Group\_Theory} is built, it can be added at the very end of the file \texttt{GroupTheoryDictionnary.py} to \texttt{FullGroupTheory}, which is a list gathering every \texttt{GroupTheory}. The dictionary will then automatically call another file whose role is to check that any \texttt{Group\_Theory} is well constructed. More precisely, it checks that the number of irreducible representations is consistent witht the order of the group (namely, $\sum_{\text{irrep }\rho}\dim(\rho)^2=|G|$),  that every $\rho(S)$ and $\rho(T)$ are square matrices of the correct dimensions, that $\rho(T)$ is diagonal and that they satisfy the relations $\rho(S^4)=\rho(T^N)=\rho((ST)^3)=\text{Id}$. It also checks that there are as many conjugacy classes as irreducible representations, that their sizes are compatible with the order of the group ($\sum_{\text{conjugacy class }c}|c|=|G|$) and that the character table is orthogonal (see Appendix \ref{sec:CharacterTheory}).

\texttt{Group\_Theory} suffices to determine the minimal set of a given group and compute its VVMFs. However, to get the higher-weight VVMFs, one needs the Clebsch-Gordan (CG) coefficients to perform the decomposition of the tensor products along irreducible representations. Those coefficients are collected in the folder \texttt{ClebschGordanLibrary}. Each group has its dedicated file \texttt{GroupName\_CG.py} in this folder which gives the decomposition of all tensor products of irreducible representations of this group. In order to study a new group, in parallel to the addition of a new \texttt{Group\_Theory} in \texttt{GroupTheoryDictionnary.py}, its CG coefficients must be provided in a new file in \texttt{ClebschGordanLibrary}. The constraints on the syntax, exemplified by the CG coefficients already tabulated, are as follows: we demand that \texttt{GroupName\_CG.py} contains a function \texttt{ProductGroupName}, whose arguments are two multiplets, i.e. two lists whose first entry is the name of the irreducible representation of the multiplet and the second is the list of its components, and whose output is a list of multiplets corresponding to the decomposition of the two arguments along the appropriate irreducible representations. Since the CG coefficients are defined up to an overall constant, we recommend to choose it so that no fraction appears in the CG coefficients. Only then can we guarantee that taking the tensor products interferes consistently with our normalization procedure of the VVMFs and does not produce errors when the code eliminates non-linearly independent ones.

Once \texttt{Group\_Theory} and the CG coefficients are known, everything is available to get the VVMFs of the group. The main file for this is \texttt{VVMFBuilder.py}. The user first specifies which group they are studying and the weight for which the VVMFs shall be constructed. For this, the first three lines are to be modified: the first one selects the corresponding \texttt{Group\_Theory} in the list \texttt{FullGroupTheory}, the second specifies which file is to be used in \texttt{ClebschGordanLibrary} to get the CG coefficients, while the third sets the weight. Then, the fourth line sets a Boolean called \texttt{FirstUse}, equal to \texttt{True} by default. We will specify later when it can be set to \texttt{False} in order to avoid performing the same computations several times. 

\texttt{VVMFBuilder.py} can now be executed. First, it will describe the minimal set in the Python console. Then it asks for the coefficients of the $q$-expansions of the VVMFs of the minimal set, which are obtained by a companion Mathematica program. The Python output lists the arguments to copy and indicates in which Mathematica file they should be pasted, which the user does by hand. There is one Mathematica file per case mentioned in \ref{sec:BasisVVMF}, namely one for each dimension up to 3 and two for dimension 4 according to whether the representation is cyclic. Once the arguments are given to Mathematica, it will produce the sought-after coefficients of the $q$-expansion, which should be copied and pasted back in the Python console. Then the Python program proceeds to compute the undetermined constants in \eqref{Dim2VVMFformula} or \eqref{Dim3VVMFformula} and checks that the result satisfies the expected modular transformation properties to be called a VVMF. This manipulation is repeated for every VVMF of the minimal set. For VVMFs transforming in representations for which no formula is known, i.e. for which we do not provide any Mathematica notebook, either the user finds the $q$-expansion elsewhere, or they simply provide an empty answer to the code, which accepts that a representation is skipped and will still perform the tensor products of the other VVMFs of the minimal set. If the user looks for VVMFs of a higher weight than those of the minimal set, the program will automatically perform the appropriate tensor products, eliminate the linearly-dependent VVMFs and normalize the remaining ones. The $q$-expansions of the VVMFs are then displayed in the console. At each weight, the code checks that enough linearly-independent VVMFs have been constructed, using \eqref{eq:dimMKrho}, and warns the user otherwise. If the user did not provide the $q$-expansions of some VVFMs in the minimal set, they shall then ignore potential warnings. To get the VVMFs for another weight, the user changes the weight in the last line of \texttt{VVMFBuilder.py} and executes the file once again. However, since the VVMFs of the minimal set are already known to the program, the Boolean \texttt{FirstUse} can now be set to \texttt{False} so as to skip the first steps: the program will directly perform the tensor products and give the result. If the user closes the console and reuses the file later or decides to study another finite modular group, they must ensure that \texttt{FirstUse} is \texttt{True} at least for the first launch.

\section{Benchmark model}
\label{sec:Benchmark model}
 Our algorithmic construction enables us to get the $q$-expansions of the VVMFs for any finite modular group, and to build a particle physics model based on them. In this section, we apply it to a finite modular group which, to our knowledge, has not yet been explored by the flavor physics literature:  $A_4\times\mathbb{Z}_4$ (see \cite{Victor} for a physical motivation to study this group from the point of view of flavor hierarchies).

The group $A_4\times\mathbb{Z}_4$ has level $N=12$ and has 16 irreducible representations: 12 singlets $\mathbf{1}_k,\mathbf{1}_k',\mathbf{1}_k''$ for $k=0,\dots,3$ and 4 triplets $\mathbf{3}_k$ for $k=0,\dots,3$. In the $T-$diagonal basis, the images of the generators are:
\begin{align*}
    &\mathbf{1}_k:\quad\rho(S)=(-i)^k&\rho(T)&=i^k\\
    &\mathbf{1}_k':\quad\rho(S)=(-i)^k&\rho(T)&=\omega i^k\\
    &\mathbf{1}_k'':\quad\rho(S)=(-i)^k&\rho(T)&=\omega^2 i^k\\
    &\mathbf{3}_k:\quad\rho(S)=\frac{(-i)^k}{3}
    \begin{pmatrix}
        -1&2&2\\
        2&-1&2\\
        2&2&-1
    \end{pmatrix}
    &\rho(T)&=i^k
    \begin{pmatrix}
        1&0&0\\
        0&\omega&0\\
        0&0&\omega^2
    \end{pmatrix}
\end{align*}
One easily checks that $\rho(T)^{12}=\text{Id}$ for every irreducible representation. The code tells us that the minimal set of $A_4\times\mathbb{Z}_4$ contains VVMFs of weight-1 in the representation $\mathbf{1}_3'$ and of weight-2 in the representations $\mathbf{1}_2''$ and $\mathbf{3}_0$. Indeed, for each irreducible representation, the minimal weights of the VVMFs are computed thanks to Section \ref{sec:BasisVVMF} and listed in Table~\ref{table:minWeightA4Z4}.
\begin{table}[h!]
    \centering
    \resizebox{\textwidth}{!}{
    \begin{tabular}{|c|c|c|c|c|c|c|c|c|c|c|c|c|c|c|c|c|}
        \hline
         & $\mathbf{1}_0$ & $\mathbf{1}_0'$ & $\mathbf{1}_0''$ & $\mathbf{1}_1$ & $\mathbf{1}_1'$ & $\mathbf{1}_1''$ & $\mathbf{1}_2$ & $\mathbf{1}_2'$ & $\mathbf{1}_2''$ & $\mathbf{1}_3$ & $\mathbf{1}_3'$ & $\mathbf{1}_3''$ & $\mathbf{3}_0$ & $\mathbf{3}_1$ & $\mathbf{3}_2$ & $\mathbf{3}_3$\\
         \hline
         $L$ & $0$ & $\frac{1}{3}$ & $\frac{2}{3}$ & $\frac{1}{4}$ & $\frac{7}{12}$ & $\frac{11}{12}$ & $\frac{1}{2}$ & $\frac{5}{6}$ & $\frac{1}{6}$ & $\frac{3}{4}$ & $\frac{1}{12}$ & $\frac{5}{12}$ & $\left[0,\frac{1}{3},\frac{2}{3}\right]$ & $\left[\frac{1}{4},\frac{7}{12},\frac{11}{12}\right]$ & $\left[\frac{1}{2},\frac{5}{6},\frac{1}{6}\right]$ & $\left[\frac{3}{4},\frac{1}{12},\frac{5}{12}\right]$\\
        \hline  
        $k_0$ & $0$ & $4$ & $8$ & $3$ & $7$ & $11$ & $6$ & $10$ & $2$ & $9$ & $1$ & $5$ & $2$ & $5$ & $4$ & $3$\\
        \hline
    \end{tabular}
    }
\caption{$L$  (see \eqref{eq:diagT})  and minimal weight of every irreducible representation of $A_4\times\mathbb{Z}_4$}. 
\label{table:minWeightA4Z4}
\end{table}

Only the irreducible representation $\mathbf{1}_3'$ has a VVMF of weight $1$. However, since $\chi_{\mathbf{1}_3'}(S^3T^3)=\chi_{\mathbf{1}_3'}(e)=1$, this irreducible representation is not faithful and thus cannot generate all the other ones by tensor products. It is for instance obvious that it cannot generate any of the triplets. Hence, weight-1 VVMFs are not enough to build the higher-weight ones. At weight $2$, there are two irreducible representations: $\mathbf{1}_2''$ and $\mathbf{3}_0$. The character of the representation $\mathbf{1}_3'\oplus\mathbf{1}_2''\oplus\mathbf{3}_0$ is the sum of the three characters:
 \begin{equation*}
\begin{split}
    \chi_{\mathbf{1}_3'\oplus\mathbf{1}_2''\oplus\mathbf{3}_0}=(5,2-i,-1,3,1,-\omega^2-i\omega,2+i,-2-i,-1,-\sqrt{3}i,-1,-i\omega^2-\omega,\\i\omega-\omega^2,-2+i,\sqrt{3}i,i\omega^2-\omega)
\end{split}
 \end{equation*}
 No other entry than that associated to $e$ is equal to $\dim(\mathbf{1}_3'\oplus\mathbf{1}_2''\oplus\mathbf{3}_0)=5$, so the representation is faithful and can generate every other irreducible representation by tensor products. The minimal set is thus composed of VVMFs of weight 1 and 2. Using the expression in terms of $\eta$ and the generalized hypergeometric series given in Section~\ref{sec:BasisVVMF}, the $q$-expansions, up to order $o\left(q^9\right)$, of these VVMFs are found to be:
 \begin{align*}     
     Y_{\mathbf{1}_3'}^{(1)}(\tau)&=q^{\frac{1}{12}}(1-2q-q^2+2q^3+q^4+2q^5-2q^6-2q^8)\\
    Y_{\mathbf{1}_2''}^{(2)}(\tau)&=q^{\frac{1}{6}}(1-4q+2q^2+8q^3-5q^4-4q^5-10q^6+8q^7+9q^8)\\
    Y_{\mathbf{3}_0}^{(2)}(\tau)&=
    \begin{pmatrix}
    \begin{aligned}
        &\phantom{--}1+12q+36q^2+12q^3+84q^4+72q^5+36q^6+96q^7+180q^8\\
        &-6q^{\frac{1}{3}}(1+7q+8q^2+18q^3+14q^4+31q^5+20q^6+36q^7+31q^8)\\
        &-18q^{\frac{2}{3}}(1+2q+5q^2+4q^3+8q^4+6q^5+14q^6+8q^7+14q^8)
    \end{aligned}
    \end{pmatrix}
 \end{align*}
For higher weights, VVMFs are obtained from tensor products of the ones above. Thanks to our code, we checked that tensor products of those VVMFs generate every linearly-independent VVMF up to weight 10 and got their $q$-expansions that we report in Appendix~\ref{sec:Higher weight VVMFs}.

Now that we have the VVMFs of $A_4\times\mathbb{Z}_4$, we are able to study a (supersymmetric) CP-symmetric \cite{Novichkov:2019sqv} model of lepton flavor based on this group. Consider the following representations:
$$L\sim\mathbf{1}_1'\oplus\mathbf{1}_1'\oplus\mathbf{1}_3'' \ , \quad E^c\sim\mathbf{1}_1\oplus\mathbf{1}_3'\oplus\mathbf{1}_3' \ ,$$
respectively associated with the weights $(k_{L_1},k_{L_2},k_{L_3})$ and $(k_{E^c_1},k_{E^c_2},k_{E^c_3})$. Higgs fields are taken to be modular-invariant, i.e. to have vanishing weights and to transform as $\mathbf{1}_0$. Moreover, we assume that neutrino masses are generated by the dimension-5 Weinberg operator. With these choices, the mass matrices are:
$$M_e=
\begin{pmatrix}
    \alpha_1^iY_{\mathbf{1}_2'',i}^{k_{L_1}+k_{E^c_1}} & \alpha_4^iY_{\mathbf{1}_2'',i}^{k_{L_2}+k_{E^c_1}} &\alpha_7^iY_{\mathbf{1}_0',i}^{k_{L_3}+k_{E^c_1}}\\
    \alpha_2^iY_{\mathbf{1}_0',i}^{k_{L_1}+k_{E^c_2}} & \alpha_5^iY_{\mathbf{1}_0',i}^{k_{L_2}+k_{E^c_2}} &\alpha_8^iY_{\mathbf{1}_2,i}^{k_{L_3}+k_{E^c_2}}\\
    \alpha_3^iY_{\mathbf{1}_0,i}^{k_{L_1}+k_{E^c_3}} & \alpha_6^iY_{\mathbf{1}_0,i}^{k_{L_2}+k_{E^c_3}} &\alpha_9^iY_{\mathbf{1}_2'',i}^{k_{L_3}+k_{E^c_3}}\\
\end{pmatrix}v_d \ ,$$
$$M_\nu = 
\begin{pmatrix}
    g_1^iY_{\mathbf{1}_2',i}^{2k_{L_1}} & g_4^iY_{\mathbf{1}_2',i}^{k_{L_1}+k_{L_2}} & g_5^iY_{\mathbf{1}_0,i}^{k_{L_1}+k_{L_3}}\\
    g_4^iY_{\mathbf{1}_2',i}^{k_{L_1}+k_{L_2}} & g_2^iY_{\mathbf{1}_2',i}^{2k_{L_2}} & g_6^iY_{\mathbf{1}_0,i}^{k_{L_2}+k_{L_3}}\\
    g_5^iY_{\mathbf{1}_0,i}^{k_{L_1}+k_{L_3}} & g_6^iY_{\mathbf{1}_0,i}^{k_{L_2}+k_{L_3}} & g_3^iY_{\mathbf{1}_2'',i}^{2k_{L_3}}
\end{pmatrix}\frac{v_u^2}{\Lambda} \ ,$$
where indices $i$ indicate that there are potentially several VVMFs transforming in a given representations. $v_{d,u}$ are the vacuum expectation values of the appropriate Higgs fields, $\Lambda$ is the new physics scale which generates the Weinberg operator, and $\alpha_n^i,g_n^i$ are free parameters of the model. CP-symmetry restricts the free parameters to be real.

These matrices appear in the basis where the Kähler potential is as in \eqref{eq:Kahler}. To extract the phenomenological predictions of the model, one must first redefine the superfields so as to make their kinetic terms canonical. With the choice of Kähler potential made in this paper, this means that the superfields become:
$$\Phi_i\rightarrow(2\text{Im}(\tau))^{\frac{k_I}{2}}\Phi_I \ .$$
The observables of interest are the charged lepton masses, the differences of squared neutrino masses, the mixing angles and the phase of the PMNS matrix. Their values are collected in Table~\ref{LeptonicObservables}.
\begin{table}[h]
    \centering
    \begin{tabular}{|c|c|c|c|c|c|c|c|}
    \hline
    Observables & $\frac{m_e}{m_\mu}$ & $\frac{m_\mu}{m_\tau}$ & $r\equiv\frac{\Delta m_{21}^2}{\Delta m_{31}^2}$ & $\sin^2(\theta_{12})$ & $\sin^2(\theta_{13})$ & $\sin^2(\theta_{23})$ & $\delta_{\text{CP}}$\\
    \hline
    Central Values & $0.00474$ & $0.0588$ & $0.02956$ & $0.303$ & $0.02225$ & $0.451$ & $1.29\pi$\\
    \hline
    1$\sigma$ error & $0.00004$ & $0.0005$ & $0.00084$ & $0.012$ & $0.00059$ & $0.019$ & $0.2\pi$\\
    \hline
    \end{tabular}
    \caption{Leptonic observables values: the ratio of neutrino squared masses difference, the mixing angles and the CP-violation phases come from NuFIT 5.2 \cite{nufit52,Esteban:2020cvm} while the charged leptons masses are taken at the GUT scale \cite{Antusch:2013jca}. See \cite{Ding:2023ydy} for a discussion on the running of the observables.}
    \label{LeptonicObservables}
\end{table}

Since those are only dimensionless observables, we can factor a free parameter in the mass matrices, namely $\alpha_1^1(2\text{Im}(\tau))^{\frac{k_{L_1}+k_{E^c_1}}{2}}$ for the charged lepton mass matrix and $g_1^1(2\text{Im}(\tau))^{k_{L_1}}$ for the neutrino mass matrix. This allows us to fix the values of $m_\tau=(1.293\pm0.007)\text{GeV}$ and $\Delta m_{21}^2 = (7.41\pm 0.21)\times10^{-5}\text{eV}^2$ by hand through an appropriate choice of the scale factors $\alpha_1^1v_d$ and $g_1^1\frac{v_u^2}{\Lambda}$. Eventually, the canonical mass matrices that contain all the physical parameters are:
$$M_e^{\text{can}}=
\begin{pmatrix}
    \hat{\alpha}_1^iY_{\mathbf{1}_2'',i}^{k_{L_1}+k_{E^c_1}} & \hat{\alpha}_4^iy^{\frac{k_{21}^e-k_{11}^e}{2}}Y_{\mathbf{1}_2'',i}^{k_{L_2}+k_{E^c_1}} &\hat{\alpha}_7^iy^{\frac{k_{31}^e-k_{11}^e}{2}}Y_{\mathbf{1}_0',i}^{k_{L_3}+k_{E^c_1}}\\
    \hat{\alpha}_2^iy^{\frac{k_{12}^e-k_{11}^e}{2}}Y_{\mathbf{1}_0',i}^{k_{L_1}+k_{E^c_2}} & \hat{\alpha}_5^iy^{\frac{k_{22}^e-k_{11}^e}{2}}Y_{\mathbf{1}_0',i}^{k_{L_2}+k_{E^c_2}} &\hat{\alpha}_8^iy^{\frac{k_{32}^e-k_{11}^e}{2}}Y_{\mathbf{1}_2,i}^{k_{L_3}+k_{E^c_2}}\\
    \hat{\alpha}_3^iy^{\frac{k_{13}^e-k_{11}^e}{2}}Y_{\mathbf{1}_0,i}^{k_{L_1}+k_{E^c_3}} & \hat{\alpha}_6^iy^{\frac{k_{23}^e-k_{11}^e}{2}}Y_{\mathbf{1}_0,i}^{k_{L_2}+k_{E^c_3}} &\hat{\alpha}_9^iy^{\frac{k_{33}^e-k_{11}^e}{2}}Y_{\mathbf{1}_2'',i}^{k_{L_3}+k_{E^c_3}}\\
\end{pmatrix}\alpha_1^1y^{\frac{k_{11}^e}{2}}v_d$$

$$M_\nu^{\text{can}} = 
\begin{pmatrix}
    \hat{g}_1^iY_{\mathbf{1}_2',i}^{2k_{L_1}} & \hat{g}_4^iy^{\frac{k_{21}^\nu-k_{11}^\nu}{2}}Y_{1_2',i}^{k_{L_1}+k_{L_2}} & \hat{g}_5^iy^{\frac{k_{31}^\nu-k_{11}^\nu}{2}}Y_{\mathbf{1}_0,i}^{k_{L_1}+k_{L_3}}\\
    \hat{g}_4^iy^{\frac{k_{21}^\nu-k_{11}^\nu}{2}}Y_{1_2',i}^{k_{L_1}+k_{L_2}} & \hat{g}_2^iy^{\frac{k_{22}^\nu-k_{11}^\nu}{2}}Y_{\mathbf{1}_2',i}^{2k_{L_2}} & \hat{g}_6^iy^{\frac{k_{32}^\nu-k_{11}^\nu}{2}}Y_{\mathbf{1}_0,i}^{k_{L_2}+k_{L_3}}\\
    \hat{g}_5^iy^{\frac{k_{31}^\nu-k_{11}^\nu}{2}}Y_{\mathbf{1}_0,i}^{k_{L_1}+k_{L_3}} & \hat{g}_6^iy^{\frac{k_{32}^\nu-k_{11}^\nu}{2}}Y_{\mathbf{1}_0,i}^{k_{L_2}+k_{L_3}} & \hat{g}_3^iy^{\frac{k_{33}^\nu-k_{11}^\nu}{2}}Y_{\mathbf{1}_2'',i}^{2k_{L_3}}
\end{pmatrix}\frac{g_1^1y^{\frac{k_{11}^{\nu}}{2}}v_u^2}{\Lambda}$$
where we used the notations $\hat{\alpha}_j^i=\frac{\alpha_j^i}{\alpha_1^1}$, $\hat{g}_j^i=\frac{g_j^i}{g_1^1}$, $y=2\text{Im}(\tau)$, $k_{ij}^e=k_{L_i}+k_{E^c_j}$ and $k_{ij}^\nu=k_{L_i}+k_{L_j}$. These canonical mass matrices predict the values of all observables, given the values of the free parameters and the modulus $\tau$. In other words, for each observable there is a function $O_i(\tau,\hat{\alpha}_j^i,\hat{g}_j^i)$ which yields the value of the $i$-th observable. The parameters of the model are then determined by fitting the predicted values of the observables to the measured ones, using a least squares method: we minimize the function
$$\chi^2(\tau,\hat{\alpha}_j^i,\hat{g}_j^i)=\sum_i\left(\frac{O_i(\tau,\hat{\alpha}_j^i,\hat{g}_j^i)-O_{i,\text{measured}}}{\sigma_i}\right)^2$$
where $O_{i,\text{measured}}$ is the measured value of the $i$-th observable and $\sigma_i$ is the incertitude of this measure. Scanning over the parameter space, we find the points that make the predicted observables closest to the measured ones.

Among the possible choices for the weights of the lepton superfields, some are able to reproduce very precisely the data with a few number of parameters. For example, the choice $(k_{L_1},k_{L_2},k_{L_3},k_{E_1},k_{E_2},k_{E_3})=(3,5,1,1,1,5)$  admits seven parameters in addition to the two real numbers corresponding to $\tau$. For this choice of weights, there only exists one VVMF per representation and weight, so we drop the exponent $i$ of the free parameters. Note that by definition $\hat{\alpha}_1=\hat{g}_1=1$, hence they are not displayed in the result of the fit. The point in the parameter space that minimizes the $\chi^2$ function ($\chi^2=0.00531$) is:
\begin{align*}
\tau =& \phantom{-}0.35478+1.82352i, &\hat{\alpha}_2 &= -0.518,&\hat{\alpha}_3 &= -9.81519,\\
{\hat{\alpha}}_4 =& -0.49369, &\hat{\alpha}_ 5 &= 4.37212,&\hat{g}
_2&=2.62511,\\
\hat{g}_3=&\phantom{-}0.3702,&\hat{g}_4&=-1.02808
\end{align*}
while the scale factors are:
$$\alpha_1v_d=0.00104\text{ GeV},\qquad\qquad \frac{g_1v_u^2}{\Lambda}=4.6872\text{ meV.}$$
It leads to the following predictions for the masses:
\[
\renewcommand{\arraystretch}{1.3}
\begin{array}{r@{{}={}}l@{\qquad \quad }r@{{}={}}l@{\qquad \quad }r@{{}={}}l}
\dfrac{m_e}{m_\mu} & 0.00474, &
\dfrac{m_\mu}{m_\tau} & 0.0588, &
\dfrac{\Delta m^2_{21}}{\Delta m^2_{31}} & 0.02956, \\
m_1 & 50.07~\text{meV}, & m_2 & 50.80~\text{meV}, & m_3 & 0.12~\text{meV}, \\
m_{\beta\beta} & 26.31~\text{meV}, & \sum_i m_i & 101~\text{meV} \ ,
\end{array}
\]
and for the mixing angles and phases:
\[
\renewcommand{\arraystretch}{1.3}
\begin{array}{r@{{}={}}l@{\qquad \quad }r@{{}={}}l@{\qquad \quad }r@{{}={}}l}
\sin^2\theta_{12} & 0.303, & \sin^2\theta_{13} & 0.0222, & \sin^2\theta_{23} & 0.451, \\
\delta_{\text{CP}} & 1.275\pi, & \alpha_{21} & 1.276\pi, & \alpha_{31} & 1.705\pi \ .
\end{array}
\]
All the observables predicted lie in the experimental $1\sigma$ range. This shows that this model based on $A_4\times\mathbb{Z}_4$ is able to correctly reproduce the experimental values to a good precision  with few parameters and to generate a CP-phase by spontaneous breaking.

\section{Conclusion}
Building a modular-invariant supersymmetric field theory begins by the choice of a finite modular group, which describes how the fields other than the modulus $\tau$ transform. For the action of the theory to be invariant, the couplings of these fields shall often transform under the action of the finite modular group, in which case they are functions of $\tau$ known as Vector-Valued Modular Forms. While considering new finite modular groups permits to build new models, being able to compute the values of the models' couplings and thus its predictions demands that one evaluates the associated VVMFs for a given vacuum expectation value of the modulus. Consequently, the study of a new finite modular group only becomes possible when its VVMFs are known.

In this paper, we presented an algorithmic procedure, as well as its Python-Mathematica implementation~\cite{GitHub_vsaulquin}, that combined the strengths of two approaches. For homogeneous finite groups $\Gamma_N'$, one constructs any VVMF as a tensor product of a handful of low-weight ones, that can be found using tractable techniques \cite{Feruglio:2017spp}. For other finite modular groups, provided one considers VVMFs in low-dimensional representations, one can rely on the formulae of Ref.\cite{Liu:2021gwa}, which are based on generalized hypergeometric functions or recursive methods. These formulae yield the lowest-weight VVMF of a given representation, hence they ought to be combined with an approach based on tensor products. One then faces the question: what is the minimal set of VVMFs that shall be constructed using Ref.~\cite{Liu:2021gwa}, before one can focus on their tensor products? Explicit examples, in particular that of the group $2O$, highlight that the answer cannot simply reduce to the lowest-weight VVMFs, which would suffice e.g. for $\Gamma_3$ \cite{Feruglio:2017spp}. One obvious but tedious fail-and-retry procedure consists in building the lowest-weight VVMFs, performing their tensor products, checking if some VVMF is missing at some higher weight, if so iterating with the next-to-lowest VVMFs, etc. Instead, we proposed a criterion that directly applies to a set of VVMFs and determines whether it has the potential to generate every VVMF of the group without having to perform its tensor products explicitly.
Although VVMFs are characterized by a weight and a representation, explicit examples for several finite modular groups suggest that the appropriate criterion can be phrased purely in representation-theoretic terms: do the tensor products of the representations in which the VVMFs of the minimal set transform generate all other representations of the group? Then, by the Burnside-Brauer theorem, this statement is equivalent to the faithfulness of the representations in the set, which can be easily checked for finite groups once the character table is known. That this criterion on the minimal set is necessary is obvious since, given any representation, there always exists a weight at which there exists a VVMF transforming according to that representation, that it appears to be sufficient in all examples which we tested is more surprising, and justifies why we take it as the central piece of our algorithm. The procedure that determines the minimal set thus reads
: for a given finite modular group, list the representations in which the lowest-weight VVMF transform using e.g. Eq.~\eqref{eq:dimMKrho}, test using the character table of the group whether they form a faithful representation, if so stop, otherwise iterate including the next-to-lowest-weight VVMFs, and so on. 

The full procedure is implementable, which we do in a code that takes as inputs the finite modular group of interest to the user (namely, its irreducible representations, its character table and its Clebsch-Gordan coefficients) and the weight up to which the VVMFs shall be built.  We detailed in the text the data and the syntax that is expected of someone who would want to study a new finite group beyond those which we already included. Then, as we described in Section~\ref{sec:CodeStructure}, the code acts in three steps. First, it applies our criterion to determine what the candidate minimal set of the group is, i.e. up to which weight the VVMFs shall be constructed using explicit formulae. Second, it extracts the $q$-expansions of the VVMFs contained in the minimal set from the formulae given in Ref.~\cite{Liu:2021gwa}. Finally, it performs their tensor products to compute all VVMFs up to the weight specified by the user. 
Our procedure relies on one hypothesis and uses partial results from the theory of VVMFs, so we implemented in the code several sanity checks and the associated warnings to the user. The main one verifies that the number of VVMFs produced for a given representation and weight matches the theoretical number given by \eqref{eq:dimMKrho}. Indeed, our criterion to determine the minimal set 
is necessary but not sufficient, although the modular structure seems to make it sufficient for the 26 groups which we tested. We extensively listed potential caveats in the text,
which may be prevented by a mathematical result (or realized by an explicit example) that we are not aware of. Moreover, an important step of the algorithm consists in identifying the representations for which a VVMF exists at a given weight. For weights strictly larger than one, the formula \eqref{eq:dimMKrho} applies, but when weight-one VVMFs are allowed, we 
rely on formulae that only apply for representations of dimension up to five. Hence, if the finite modular group has a higher-dimensional VVMF of weight one, the code would not be able to detect it.
The code then warns the user when the group has a high-dimensional odd irreducible representation for which a weight-1 VVMF could have been missed. 
The last point is the ability to build the VVMFs of the minimal set. This part of the code relies on Ref.~\cite{Liu:2021gwa}, whose formulae or recursive methods are limited to dimensions smaller than four. For higher-dimensional representations, the user has to find the coefficients of the $q$-expansions of the desired VVMFs by another way, or ignore them.

Finally, in order to illustrate how our code helps to study field theories based on any finite modular group, that can only be explored after the associated modular forms have been found, we built an explicit model of lepton flavor based on the unexplored group $A_4\times\mathbb{Z}_4$. As usual in this context, the observables of the leptonic sector are functions of the modulus and few free parameters, connected to the few VVMFs featured in the model which our code produced. By a scan over its parameters, we showed that this model is able to precisely reproduce the leptonic mixing angles and masses, while generating the $CP$-phase from spontaneous CP breaking, with a small number of parameters. Therefore, $A_4\times\mathbb{Z}_4$, like the other exotic groups $2D_3,A_4\times\mathbb{Z}_2,2O$ or $GL(2,3)$ already explored in the literature, are as predictive as the more common $\Gamma_N'$ groups and are worth studying. Looking back at Table~\ref{Finite modular groups} of the first finite modular groups, one sees that many of them remain to be explored, in particular the groups made of products and semi-direct products of the cyclic groups $\mathbb{Z}_N$. Their study could lead to interesting phenomenological behaviors.
\acknowledgments
We thank Simon Beaudoin for his comments regarding the user experience of the code, and Emilian Dudas for valuable discussions. VS is supported by the Interdisciplinary Thematic Institute QMat, as part of the ITI 2021-2028 program of the University of Strasbourg, CNRS and Inserm, and were supported by IdEx Unistra (ANR-10-IDEX-0002), SFRI STRAT' US project (ANR-20-SFRI-0012), and EUR QMAT (ANR-17-EURE-0024). These funding awards were made within the framework of the French Investments for the Future Program. QB is supported by IdEx funding ``Attractivity 2025'' of the University of Strasbourg.

\appendix
\section{Basic facts about character theory}
\label{sec:CharacterTheory}
We recall here some basic facts about the character theory of finite groups. Let $G$ be a finite group, and $\rho$ be an (irreducible) representation of $G$. Then the (irreducible) character of $\rho$ is a function of $G$
to $\mathbb C$ defined by:
$$\chi_\rho(g)=\text{Tr}(\rho(g))$$
Since $\rho$ is a group representation and by cyclicity of the trace, the characters are class functions, namely they are constant on the conjugacy classes: for every $g,h\in G$,
$$\chi_\rho(hgh^{-1})= \chi_\rho(g)$$
The canonical scalar product on the characters is as follows: for two representations $\rho_1,\rho_2$ of $G$,
\begin{equation}
\label{ScalarProductCharacters}
\big<\chi_{\rho_1},\chi_{\rho_2}\big>=\frac{1}{|G|}\sum_{g\in G}\chi_{\rho_1}(g)\overline{\chi_{\rho_2}(g)}
\end{equation}
One can show that the family of irreducible characters is orthonormal for this scalar product. An even stronger theorem states that the irreducible characters form a basis of the class functions. A consequence of this is that there are as many irreducible characters as conjugacy classes of $G$. Thus, the number of irreducible representations of $G$ is equal to the number of conjugacy classes of $G$. The character table is a simple way to show the character of a group. This is a table where a line corresponds to an irreducible representation and each column to a conjugacy class of $G$. The entry $i,j$ is the value of the $i$-th character on the $j$-th conjugacy class. By the discussion above, the character table must be a square and each line are orthogonal for the scalar product \eqref{ScalarProductCharacters}.

Let us mention two simple facts. First, the character evaluation on the neutral element of the group is equal to the dimension of the representation: $\chi_{\rho}(e)=\dim(\rho)$. Second, if $\rho_1,\rho_2$ are two representations of $G$ then the character of the sum simply is the sum of the characters: $\chi_{\rho_1\oplus\rho_2}=\chi_{\rho_1}+\chi_{\rho_2}$.

Finally the main property used in this paper is that a representation is faithful if and only if for all $g\neq e,\chi(g)\neq\chi(e)$. Indeed, one can show that:
$$\ker(\rho)=\{g\in G,\chi_\rho(g)=\chi_\rho(e)\}$$
By definition, a representation is faithful when its kernel is reduced to the neutral element so the result.

\section{Some useful functions}
\label{sec:Useful functions}
We define in this appendix some functions appearing in the analytic expressions of VVMFs. The first one is the Dedekind eta function $\eta$:
$$\eta(\tau)=q^{\frac{1}{24}}\prod_{k=1}^\infty\left(1-q^k\right)$$
with $q=e^{2i\pi\tau}$. (Let us stress in passing that the common notation $q^{\frac{1}{24}}$, leaves the modular transformations of the object somewhat ambiguous: $q^{\frac{1}{24}}\xrightarrow[]{T} e^{2i\pi/24}q^{\frac{1}{24}}$. The same applies to $q^{r_i}$ introduced in Section~\ref{sec:VVMF}.)
This function is holomorphic and satisfies:
$$\eta(\tau+1)=e^{2i\pi/24}\eta(\tau)\qquad\text{and}\qquad\eta\left(-\frac{1}{\tau}\right)=\sqrt{-i\tau}\eta(\tau)$$
Note that VVMFs transforming in dimension-1 representations can be formed from $\eta$. For instance, $\eta^2$ transforms as a VVMF of weight $1$ such that $\rho(T)=e^{i\frac{\pi}{6}},\rho(S)=i$.

The local parameter involved in the explicit expressions of the basis of the VVMF modules in \ref{sec:BasisVVMF} is:
$$K(\tau)=\frac{E_4(\tau)^3-E_6(\tau)^2}{E_4(\tau)^3}$$
The generalized hypergeometric series are defined as:
$${}_{p}F_q(x_1,\dots,x_p;y_1,\dots,y_q;z)=\sum_{n=0}^\infty\frac{\prod_{j=1}^px_j^{(m)}}{\prod_{j=1}^qy_j^{(m)}}\frac{z^m}{m!}$$
where the rising factorial (or Pochhammer symbol) is used:
$$x^{(m)}=\prod_{k=1}^m(x+k-1).$$
Note that when $p\leq q$ the series converges on the whole complex space, when $p=q+1$ it converges for $|z|<1$ and when $p>q+1$ it diverges as soon as $z\neq 0$. In the context of VVMF, $p=q+1$ and $z=K(\tau)$, so since $K(\tau)\xrightarrow[\tau\rightarrow i\infty]{} 0 $ the formulae in Section~\ref{sec:BasisVVMF} are well defined in a region close to $i\infty$. $q$-expansions of the VVMF derived in this region are then valid on the whole Poincaré half-plane.

\section{Recursive method for dimension-4 VVMFs}
\label{sec:Recursive method}
We now present the recursive method to get the coefficients of the $q$-expansion for the minimal VVMF $F_{k_0}$ of a 4-dimensional irreducible representation. First we define the zero-weight function $\tilde{F}_{k_0}=\eta^{-2k_0}F_{k_0}$. This function can be parametrized with the help of the local parameter $K$:
$$\tilde{F}_{k_0}=
\left(
\begin{matrix}
K^{r_1'}\sum_{n\geq0}a_1(n)K^n\\
K^{r_2'}\sum_{n\geq0}a_2(n)K^n\\
K^{r_3'}\sum_{n\geq0}a_3(n)K^n\\
K^{r_4'}\sum_{n\geq0}a_4(n)K^n\\
\end{matrix}
\right)
$$
where $r_i'=r_i-\frac{k_0}{12}$ and the $r_i$ have been defined in \eqref{eq:diagT}. Following the method in \cite{Ding:2023ydy} based on the result of \cite{franc2018constructionsvectorvaluedmodularforms}, we are able to get a recursive formula on the coefficients $a_i(n)$ for any dimension-4 representation. One can show that in the cyclic case the coefficients $a_i(n)$ must satisfy the recursive relation:
\begin{align*}
    0&=\left[(r_i'+n)_4+5(r_i'+n)_3+\left(a+\frac{155}{36}\right)(r_i'+n)_2+\left(\frac{5a}{6}+b+\frac{5}{18}\right)(r_i'+n)_1+c\right]a_i(n)\\
    &-\left[2(r_i'+n-1)_4+13(r_i'+n-1)_3+\left(a+\frac{160}{9}\right)(r_i'+n-1)_2\right.\\
    &\qquad\qquad\qquad\qquad\qquad\qquad\qquad\qquad\qquad\qquad\left.+\left(\frac{35}{9}+\frac{4a}{3}+b\right)(r_i'+n-1)_1\right]a_i(n-1)\\
    &+\left[(r_i'+n-2)_4+8(r_i'+n-2)_3+\frac{128}{9}(r_i'+n-2)_2+\frac{40}{9}(r_i'+n-2)_1\right]a_i(n-2)
\end{align*}
with the initial conditions $a_i(-2)=a_i(-1)=0$ and $a_i(0)=1728^{-r_i'}$ and where the falling factorial is:
$$(x)_m=\prod_{k=1}^m(x-m+1)$$
The coefficients $a,b,c$ only depend on the $r_i'$:
\begin{align*}
a&=\sigma_2(r_1',r_2',r_3',r_4')-\frac{11}{36}\\
b&=-\sigma_3(r_1',r_2',r_3',r_4')+\frac{a}{6}+\frac{1}{36}\\
c&=\sigma_4(r_1',r_2',r_3',r_4')
\end{align*}
and $\sigma_i$ are the elementary symmetric polynomials:
\begin{align*}
    \sigma_2(w,x,y,z)&=wx+wy+wz+xy+xz+yz\\
    \sigma_3(w,x,y,z)&=wxy+wxz+wyz+xyz\\
    \sigma_4(w,x,y,z)&=wxyz
\end{align*}
In the non-cyclic case, the recursive relation becomes:
\begin{align*}
    0&=\left[108(r_i'+n)_4+576(r_i'+n)_3+(555-108(a+b))(r_i'+n)_2\right.\\
    &\left.\qquad\qquad\qquad\qquad\qquad\qquad\qquad\qquad+(50-54a-90b)(r_i'+n)_1-(6a+108c)\right]a_i(n)\\
    &-\left[216(r_i'+n-1)_4+1476(r_i'+n-1)_3+(2154-108(a+b))(r_i'+n-1)_2\right.\\
    &\left.\qquad\qquad\qquad\qquad\qquad\qquad\qquad\qquad+(520-108a-144b(r_i'+n-1)_1+12a\right]a_i(n-1)\\
    &+\left[108(r_i'+n-2)_4+900(r_i'+n-2)_3+1680(r_i'+n-2)_2+560(r_i'+n-2)_1\right]a_i(n-2)
\end{align*}
and $a,b,c$ are:
\begin{align*}
a&=-3\sigma_3(r_1',r_2',r_3',r_4')+\frac{1}{2}\sigma_2(r_1',r_2',r_3',r_4')-\frac{1}{24}\\
b&=3\sigma_3(r_1',r_2',r_3',r_4')-\frac{3}{2}\sigma_2(r_1',r_2',r_3',r_4')+\frac{13}{72}\\
c&=-\sigma_4(r_1',r_2',r_3',r_4')-\frac{a}{18}
\end{align*}

\section{Character tables of relevant groups}
\label{sec:Group theory dictionnary}
In this section, we give some the irreducible representations and the character table for the groups mentioned in the main text. Note that in the character tables we use the notation $nC_k$ to denote a conjugacy class of size $n$ and order $k$.

\begin{itemize}
\item $A_4=\left<S,T|S^2=(ST)^3=S^2TS^{-2}T^{-1}=T^3=1\right>$ whose \texttt{GAPId} is $[12,3]$. It has 4 irreducible representations: $\mathbf{1}, \mathbf{1}', \mathbf{1}'', \mathbf{3}$. Its character table is given in Table~\ref{CharacterTableA4}.
\begin{table}[h]
    \centering
    \begin{tabular}{|c|c|c|c|c|}
    \hline
    & $1C_1$ & $4C_3$ & $4C_3'$ & $3C_2$ \\
    \hline
    & $e$ & $T$ & $T^2$ & $S$\\
    \hline
    $\mathbf{1}$ & $1$ & $1$ & $1$ & $1$\\
    \hline
    $\mathbf{1}'$ & $1$ & $\omega$ & $\omega^2$ & $1$\\
    \hline
    $\mathbf{1}''$ & $1$ & $\omega^2$ & $\omega$ & $1$ \\
    \hline
    $\mathbf{3}$ & $3$ & $0$ & $0$ & $-1$ \\
    \hline
    \end{tabular}
    \caption{Character table of $A_4$}
    \label{CharacterTableA4}
\end{table}

\item $T'=\left<S,T|S^4=(ST)^3=S^2TS^{-2}T^{-1}=T^3=1\right>$ whose \texttt{GAPId} is $[24,3]$. It is the double covering of the previous group and has 7 irreducible representations: $\mathbf{1}, \mathbf{1}', \mathbf{1}'', \mathbf{2}, \mathbf{2}',$ $ \mathbf{2}'', \mathbf{3}$. Its character table is given in Table~\ref{CharacterTableT'}.
\begin{table}[h!]
    \centering
    \begin{tabular}{|c|c|c|c|c|c|c|c|}
    \hline
    & $1C_1$ & $1C_2$ & $6C_4$ & $4C_6$ & $4C_3$ & $4C_3'$ & $4C_6'$\\
    \hline
    & $e$ & $S^2$ & $S$ & $S^2T$ & $T^2$ & $T$ & $S^2(ST)^2$\\
    \hline
    $\mathbf{1}$ & $1$ & $1$ & $1$ & $1$ & $1$ & $1$ & $1$\\
    \hline
    $\mathbf{1}'$ & $1$ & $1$ & $1$ & $\omega$ & $\omega^2$ & $\omega$ & $\omega^2$\\
    \hline
    $\mathbf{1}''$ & $1$ & $1$ & $1$ & $\omega^2$ & $\omega$ & $\omega^2$ & $\omega$\\
    \hline
    $\mathbf{2}$ & $2$ & $-2$ & $0$ & $\omega^2$ & $-\omega$ & $-\omega^2$ & $\omega$\\
    \hline
    $\mathbf{2}'$ & $2$ & $-2$ & $0$ & $1$ & $-1$ & $-1$ & $1$\\
    \hline
    $\mathbf{2}''$ & $2$ & $-2$ & $0$ & $\omega$ & $-\omega^2$ & $-\omega$ & $\omega^2$\\
    \hline
    $\mathbf{3}$ & $3$ & $3$ & $-1$ & $0$ & $0$ & $0$ & $0$\\
    \hline
    \end{tabular}
    \caption{Character table of $T'$}
    \label{CharacterTableT'}
\end{table}

\item $2O=\left<S,T|S^4=(ST)^3=S^2TS^{-2}T^{-1}=S^2T^4=1\right>$ whose \texttt{GAPId} is $[48,28]$. It has 8 irreducible representations: $\mathbf{1},\mathbf{1}',\mathbf{2},\hat{\mathbf{2}},\hat{\mathbf{2}}',\mathbf{3},\mathbf{3'},\hat{\mathbf{4}}$. Its character table is given in Table~\ref{CharacterTable2O}.
\begin{table}[h!]
    \centering
    \begin{tabular}{|c|c|c|c|c|c|c|c|c|}
    \hline
    & $1C_1$ & $1C_2$ & $6C_8$ & $6C_8'$ & $6C_4$ & $8C_6$ & $8C_3$ &$12C_4$\\
    \hline
    & $e$ & $S^2$ & $T$ & $ST^2$ & $(ST^2)^2$ & $S^2TS$ & $S^2(ST)^2T^2$ & $S^3T^2ST$\\
    \hline
    $\mathbf{1}$ & $1$ & $1$ & $1$ & $1$ & $1$ & $1$ & $1$ & $1$\\
    \hline
    $\mathbf{1}'$ & $1$ & $1$ & $-1$ & $-1$ & $1$ & $1$ & $1$ & $-1$\\
    \hline
    $\mathbf{2}$ & $2$ & $2$ & $0$ & $0$ & $2$ & $-1$ & $-1$ & $0$\\
    \hline
    $\hat{\mathbf{2}}$ & $2$ & $-2$ & $-\sqrt{2}$ & $\sqrt{2}$ & $0$ & $1$ & $-1$ & $0$\\
    \hline
    $\hat{\mathbf{2}}'$ & $2$ & $-2$ & $\sqrt{2}$ & $-\sqrt{2}$ & $0$ & $1$ & $-1$ & $0$\\
    \hline
    $\mathbf{3}$ & $3$ & $3$ & $1$ & $1$ & $-1$ & $0$ & $0$ & $-1$\\
    \hline
    $\mathbf{3}'$ & $3$ & $3$ & $-1$ & $-1$ & $-1$ & $0$ & $0$ & $1$\\
    \hline
    $\hat{\mathbf{4}}$ & $4$ & $-4$ & $0$ & $0$ & $0$ & $-1$ & $1$ & $0$\\
    \hline
    \end{tabular}
    \caption{Character table of $2O$}
    \label{CharacterTable2O}
\end{table}
\item $GL(2,3)=\left<S,T|S^4=(ST)^3=S^2TS^{-2}T^{-1}=T^8=ST^4ST^{-4}=1\right>$ whose \texttt{GAPId} is $[48,29]$. It has 8 irreducible representations: $\mathbf{1},\mathbf{1}',\mathbf{2},\mathbf{2}',\mathbf{2}'',\mathbf{3},\mathbf{3}',\mathbf{4}$. Its character table is given in Table~\ref{CharacterTableGL(2,3)}.
\begin{table}[h!]
    \centering
    \begin{tabular}{|c|c|c|c|c|c|c|c|c|}
    \hline
    & $1C_1$ & $1C_2$ & $6C_8$ & $6C_8'$ & $6C_4$ & $8C_6$ & $8C_3$ &$12C_2$\\
    \hline
    & $e$ & $T^4$ & $T$ & $ST^2$ & $(T^2S)^2$ & $TST^3S$ & $TST^4$ & $TST^3$\\
    \hline
    $\mathbf{1}$ & $1$ & $1$ & $1$ & $1$ & $1$ & $1$ & $1$ & $1$\\
    \hline
    $\mathbf{1}'$ & $1$ & $1$ & $-1$ & $-1$ & $1$ & $1$ & $1$ & $-1$\\
    \hline
    $\mathbf{2}$ & $2$ & $2$ & $0$ & $0$ & $2$ & $-1$ & $-1$ & $0$\\
    \hline
    $\mathbf{2}'$ & $2$ & $-2$ & $\sqrt{2}i$ & $-\sqrt{2}i$ & $0$ & $-1$ & $1$ & $0$\\
    \hline
    $\mathbf{2}''$ & $2$ & $-2$ & $-\sqrt{2}i$ & $\sqrt{2}i$ & $0$ & $-1$ & $1$ & $0$\\
    \hline
    $\mathbf{3}$ & $3$ & $3$ & $1$ & $1$ & $-1$ & $0$ & $0$ & $-1$\\
    \hline
    $\mathbf{3}'$ & $3$ & $3$ & $-1$ & $-1$ & $-1$ & $0$ & $0$ & $1$\\
    \hline
    $\mathbf{4}$ & $4$ & $-4$ & $0$ & $0$ & $0$ & $1$ & $-1$ & $0$\\
    \hline
    \end{tabular}
    \caption{Character table of $GL(2,3)$}
    \label{CharacterTableGL(2,3)}
\end{table}

\item $A_4\times\mathbb{Z}_4=\left<S,T|S^4=(ST)^3=S^2TS^{-2}T^{-1}=(ST^{-1}ST)^2=1\right>$ whose \texttt{GAPId} is [48,31]. It has 16 irreducible representations: $\mathbf{1}_k,\mathbf{1}_k',\mathbf{1}_k'',\mathbf{3}_k$ for $k=0,1,2,3$. Its character table is given in Table~\ref{CharacterTableA4Z4}.

\begin{table}[h!]
    \centering
    \resizebox{\textwidth}{!}{
    \begin{tabular}{|c|c|c|c|c|c|c|c|c|c|c|c|c|c|c|c|c|}
        \hline

        & $1C_1$ & $1C_4$ & $4C_3$ & $1C_2$ & $3C_2$ & $4C_{12}$ & $1C_4'$ & $3C_4$ & $4C_3'$ & $ 4C_6$ & $3C_2'$ & $4C_{12}'$ & $4C_{12}''$ & $3C_4'$ & $4C_6'$ & $4C_{12}'''$\\
        \hline
        & $e$ & $S^2T^3$ & $T^4$ & $S^2$ & $S^3T^3$ & $T$ & $T^3$ & $S^3$ & $S^2T^2$ & $S^2T^4$ & $ST^3$ & $S^2(ST)^2S$ & $S^2T$ & $S$ & $T^2$ & $(ST)^2S$\\
        \hline
        $\mathbf{1}_0$ & $1$ & $1$ & $1$ & $1$ & $1$ & $1$ & $1$ & $1$ & $1$ & $1$ & $1$ & $1$ & $1$ & $1$ & $1$ & $1$ \\
        \hline
        $\mathbf{1}_0'$ & $1$ & $1$ & $\omega$ & $1$ & $1$ & $\omega$ & $1$ & $1$ & $\omega^2$ & $\omega$ & $1$ & $\omega^2$ & $\omega$ & $1$ & $\omega^2$ & $\omega^2$\\
        \hline
        $\mathbf{1}_0''$ & $1$ & $1$ & $\omega^2$ & $1$ & $1$ & $\omega^2$ & $1$ & $1$ & $\omega$ & $\omega^2$ & $1$ & $\omega$ & $\omega^2$ & $1$ & $\omega$ & $\omega$ \\
        \hline 
        $\mathbf{1}_1$ & $1$ & $i$ & $1$ & $-1$ & $1$ & $i$ & $-i$ & $i$ & $1$ & $-1$ & $-1$ & $i$ & $-i$ & $-i$ & $-1$ & $-i$\\
        \hline
        $\mathbf{1}_1'$ & $1$ & $i$ & $\omega$ & $-1$ & $1$ & $i\omega$ & $-i$ & $i$ & $\omega^2$ & $-\omega$ & $-1$ & $i\omega^2$ & $-i\omega$ & $-i$ & $-\omega^2$ & $-i\omega^2$ \\
        \hline
        $\mathbf{1}_1''$ & $1$ & $i$ & $\omega^2$ & $-1$ & $1$ & $i\omega^2$ & $-i$ & $i$ & $\omega$ & $-\omega^2$ & $-1$ & $i\omega$ & $-i\omega^2$ & $-i$ & $-\omega$ & $-i\omega$ \\
        \hline
        $\mathbf{1}_2$ & $1$ & $-1$ & $1$ & $1$ & $1$ & $-1$ & $-1$ & $-1$ & $1$ & $1$ & $1$ & $-1$ & $-1$ & $-1$ & $1$ & $-1$ \\
        \hline
        $\mathbf{1}_2'$ & $1$ & $-1$ & $\omega$ & $1$ & $1$ & $-\omega$ & $-1$ & $-1$ & $\omega^2$ & $\omega$ & $1$ & $-\omega^2$ & $-\omega$ & $-1$ & $\omega^2$ & $-\omega^2$\\
        \hline
        $\mathbf{1}_2''$ & $1$ & $-1$ & $\omega^2$ & $1$ & $1$ & $-\omega^2$ & $-1$ & $-1$ & $\omega$ & $\omega^2$ & $1$ & $-\omega$ & $-\omega^2$ & $-1$ & $\omega$ & $-\omega$ \\
        \hline
        $\mathbf{1}_3$ & $1$ & $-i$ & $1$ & $-1$ & $1$ & $-i$ & $i$ & $-i$ & $1$ & $-1$ & $-1$ & $-i$ & $i$ & $i$ & $-1$ & $i$\\
        \hline
        $\mathbf{1}_3'$ & $1$ & $-i$ & $\omega$ & $-1$ & $1$ & $-i\omega$ & $i$ & $-i$ & $\omega^2$ & $-\omega$ & $-1$ & $-i\omega^2$ & $i\omega$ & $i$ & $-\omega^2$  & $i\omega^2$ \\
        \hline
        $\mathbf{1}_3''$ & $1$ & $-i$ & $\omega^2$ & $-1$ & $1$ & $-i\omega^2$ & $i$ & $-i$ & $\omega$ & $-\omega^2$ & $-1$ & $-i\omega$ & $i\omega^2$ & $i$ & $-\omega$ & $i\omega$\\
        \hline
        $\mathbf{3}_0$ & $3$ & $3$ & $0$ & $3$ & $-1$ & $0$ & $3$ & $-1$ & $0$ & $0$ & $-1$ & $0$ & $0$ & $-1$ & $0$ & $0$ \\
        \hline
        $\mathbf{3}_1$ & $3$ & $3i$ & $0$ & $-3$ & $-1$ & $0$ & $-3i$ & $-i$ & $0$ & $0$ & $1$ & $0$ & $0$ & $i$ & $0$ & $0$ \\
        \hline
        $\mathbf{3}_2$ & $3$ & $-3$ & $0$ & $3$ & $-1$ & $0$ & $-3$ & $1$ & $0$ & $0$ & $-1$ & $0$ & $0$ & $1$ & $0$ & $0$\\
        \hline
        $\mathbf{3}_3$ & $3$ & $-3i$ & $0$ & $-3$ & $-1$ & $0$ & $3i$ & $i$ & $0$ & $0$ & $1$ & $0$ & $0$ & $-i$ & $0$ & $0$\\
        \hline
        
    \end{tabular}
    }
    \caption{Character table of $A_4\times\mathbb{Z}_4$}
    \label{CharacterTableA4Z4}
\end{table}
Here are its Clebsch-Gordan coefficients: let $r_1$ and $r_2$ be two irreducible representations of $A_4\times\mathbb{Z}_4$ whose components are $r_1=(\alpha_1,\dots,\alpha_{\dim(r_1)})^T$ and $r_2=(\beta_1,\dots,\beta_{\dim(r_2)})^T$. If $r_1$ and $r_2$ both are singlets then:
$$r_1\otimes r_2=(\alpha_1\beta_1)$$
and the resulting representation is:
$$
\mathbf{1}_i\otimes\mathbf{1}_j=\mathbf{1}_{i+j},\qquad\mathbf{1}_i'\otimes\mathbf{1}_j'=\mathbf{1}_{i+j}''$$
$$\mathbf{1_i}\otimes\mathbf{1}_j'=\mathbf{1}_{i+j}',\qquad\mathbf{1}_i'\otimes\mathbf{1}_j''=\mathbf{1}_{i+j}$$
$$\mathbf{1}_i\otimes\mathbf{1}_j''=\mathbf{1}_{i+j}'',\qquad\mathbf{1}''\otimes\mathbf{1}_j''=\mathbf{1}_{i+j}'
$$
where $i+j$ msut be understood modulo 4.
If $r_1=\mathbf{1}_i,\mathbf{1}_i'$ or $\mathbf{1}_i''$ and $r_2=\mathbf{3}_j$ then $$r_1\otimes r_2=\mathbf{3}_{i+j}$$
and the coefficients are:
$$\mathbf{1}_i\otimes\mathbf{3}_j:
\begin{pmatrix}
\alpha_1\beta_1\\
\alpha_1\beta_2\\
\alpha_1\beta_3
\end{pmatrix},
\qquad
\mathbf{1}_i'\otimes\mathbf{3}_j:
\begin{pmatrix}
    \alpha_1\beta_3\\
    \alpha_1\beta_1\\
    \alpha_1\beta_2
\end{pmatrix},
\qquad
\mathbf{1}_i''\otimes\mathbf{3}_j:
\begin{pmatrix}
    \alpha_1\beta_2\\
    \alpha_1\beta_3\\
    \alpha_1\beta_1
\end{pmatrix}
$$
Finally, if $r_1=\mathbf{3}_i$ and $r_2=\mathbf{3}_j$ then the decomposition of the product is:
$$r_1\otimes r_2=\mathbf{1}_{i+j}\oplus\mathbf{1}_{i+j}'\oplus\mathbf{1}_{i+j}''\oplus\mathbf{3}_{i+j,S}\oplus\mathbf{3}_{i+j,A}$$
with the coefficients:
$$\mathbf{1}_{i+j}:\alpha_1\beta_1+\alpha_2\beta_3+\alpha_3\beta_2$$
$$\mathbf{1}_{i+j}':\alpha_3\beta_3+\alpha_1\beta_2+\alpha_2\beta_1$$
$$\mathbf{1}_{i+j}'':\alpha_2\beta_2+\alpha_1\beta_3+\alpha_3\beta_1$$
$$\mathbf{3}_{i+j,S}:
\begin{pmatrix}
    2\alpha_1\beta_1-\alpha_2\beta_3-\alpha_3\beta_2\\
    2\alpha_3\beta_3-\alpha_1\beta_2-\alpha_2\beta_1\\
    2\alpha_2\beta_2-\alpha_1\beta_3-\alpha_3\beta_1
\end{pmatrix}$$
$$\mathbf{3}_{i+j,A}:
\begin{pmatrix}
    \alpha_3\beta_2-\alpha_2\beta_3\\
    \alpha_2\beta_1-\alpha_1\beta_2\\
    \alpha_1\beta_3-\alpha_3\beta_1
\end{pmatrix}$$

\end{itemize}
\section{Higher-weight VVMFs of $A_4\times\mathbb{Z}_4$}
\label{sec:Higher weight VVMFs}
We give in this appendix the $q$-expansions of the VVMFs of $A_4\times\mathbb{Z}_4$ up to weight $10$ and order $o(q^9)$. Weight-1 and -2 VVMFs have already been reported in Section~\ref{sec:Benchmark model} as we discussed our benchmark model. The other functions used in this context can be found among the VVMFs below.
\begin{align*}
\text{Weight 3:}\\
Y_{\mathbf{1}_1}^{(3)}(\tau)&=q^{\frac{1}{4}}(1-6q+9q^2+10q^3-30q^4+11q^6+42q^7)\\
Y_{\mathbf{3}_3}^{(3)}(\tau)&=
\begin{pmatrix}
\begin{aligned}
    &-18q^{\frac{3}{4}} \left(1-6q^3+9q^6\right)\\
    &q^{\frac{1}{12}} \left(1+10q+11q^2-70q^3+49q^4-22q^5-110q^6+180q^7+130q^8\right)\\
    &-6q^{\frac{5}{12}} \left(1+5q-7q^2-3q^3-15q^4+10q^5+13q^7+33q^8\right)
\end{aligned}
\end{pmatrix}\\
\text{Weight 4:}\\
Y_{\mathbf{1}_0}^{(4)}(\tau)&=
1+240q+2160q^2+6720q^3+17520q^4+30240q^5+60480q^6+82560q^7+140400q^8\\
Y_{\mathbf{3}_0}^{(4)}(\tau)&=
\begin{pmatrix}
\begin{aligned}
    &1-84q-756q^2-2028q^3-6132q^4-10584q^5-18252q^6-28896q^7-49140q^8\\
    &6q^{\frac{1}{3}}\left(1+73q+344q^2+1134q^3+2198q^4+4681q^5+6860q^6+11988q^7+15751q^8\right)\\
    &54q^{\frac{2}{3}}\left(1+14q+65q^2+148q^3+344q^4+546q^5+1022q^6+1352q^7+2198q^8\right)
\end{aligned}
\end{pmatrix}\\
Y_{\mathbf{3}_2}^{(4)}(\tau)&=
\begin{pmatrix}
\begin{aligned}
&-6q^{\frac{1}{2}}\left(1+3q-18q^2+8q^3+9q^4+36q^5-10q^6-54q^7+18q^8\right)\\
&-18q^{\frac{5}{6}}\left(1-2q-q^2-4q^3+13q^4+8q^5-5q^6-24q^7-23q^8\right)\\
&q^{\frac{1}{6}}\left(1+8q-10q^2-100q^3+199q^4-16q^5-226q^6+452q^7-279q^8\right)
\end{aligned}
\end{pmatrix}.
\end{align*}

\begin{align*}
\text{Weight 5:}\\
Y_{\mathbf{1}_3'}^{(5)}(\tau)
&= q^{\frac{1}{12}}\left(1+238q+1679q^2+2162q^3+2401q^4-6958q^5-1442q^6-23040q^7+1918q^8\right)\\
Y_{\mathbf{1}_3''}^{(5)}(\tau)
&= q^{\frac{5}{12}}\left(1-10q+35q^2-30q^3-105q^4+238q^5-260q^7-165q^8\right)\\
Y_{\mathbf{3}_1}^{(5)}(\tau)
&=
\begin{pmatrix}
\begin{aligned}
&q^{\frac{1}{4}}\left(1+6q-27q^2-86q^3+426q^4-324q^5-589q^6+1182q^7-972q^8\right)\\
&-6q^{\frac{7}{12}}\left(1+q-25q^2+43q^3+18q^4-21q^5-89q^6-86q^7+267q^8\right)\\
&-18q^{\frac{11}{12}}\left(1-4q+2q^2+2q^3+19q^4-16q^5-49q^6+2q^7+51q^8\right)
\end{aligned}
\end{pmatrix}\\
Y_{\mathbf{3}_3}^{(5)}(\tau)
&=
\begin{pmatrix}
\begin{aligned}
&54q^{\frac{3}{4}}\left(1+12q+36q^2+6q^3+12q^4-144q^5-27q^6-300q^7+72q^8\right)\\
&q^{\frac{1}{12}}\left(1-86q-589q^2-430q^3-1487q^4+2114q^5+4066q^6+2556q^7-2942q^8\right)\\
&6q^{\frac{5}{12}}\left(1+71q+197q^2+375q^3-267q^4-86q^5-1944q^6-341q^7-1947q^8\right)
\end{aligned}
\end{pmatrix}
\end{align*}

\begin{align*}
\text{Weight 6:}\\
Y_{\mathbf{1}_0}^{(6)}(\tau)
&= 1-504q-16632q^2-122976q^3-532728q^4-1575504q^5-4058208q^6\\
&\qquad\qquad\qquad\qquad\qquad\qquad\qquad\qquad\qquad\qquad\qquad\quad-8471232q^7-17047800q^8\\
Y_{\mathbf{1}_2}^{(6)}(\tau)
&= q^{\frac{1}{2}}\left(1-12q+54q^2-88q^3-99q^4+540q^5-418q^6-648q^7+594q^8\right)\\
Y_{\mathbf{1}_2''}^{(6)}(\tau)
&= q^{\frac{1}{6}}\left(1+236q+1202q^2-1432q^3-3125q^4-10324q^5+16550q^6\right.\\
&\qquad\qquad\qquad\qquad\qquad\qquad\qquad\qquad\qquad\qquad\qquad\qquad\qquad\left.-3352q^7+38889q^8\right)\\
Y_{\mathbf{3}_0^{I}}^{(6)}(\tau)
&=
\begin{pmatrix}
\begin{aligned}
&1+144q+5724q^2+40320q^3+178452q^4+527904q^5+1339308q^6\\
&\qquad\qquad\qquad\qquad\qquad\qquad\qquad\qquad\qquad\qquad\quad+2840256q^7+5709636q^8\\
&-1458q^{\frac{1}{3}}\left(q+16q^2+98q^3+352q^4+1029q^5+2352q^6+5044q^7+9280q^8\right)\\
&-54q^{\frac{2}{3}}\left(1+80q+863q^2+4144q^3+14216q^4+36384q^5+84722q^6\right.\\
&\qquad\qquad\qquad\qquad\qquad\qquad\qquad\qquad\qquad\qquad\qquad\left.+165056q^7+314270q^8\right)\\
\end{aligned}
\end{pmatrix}\\
Y_{\mathbf{3}_0^{II}}^{(6)}(\tau)
&=
\begin{pmatrix}
\begin{aligned}
&-18q+108q^2-162q^3-72q^4-108q^5+972q^6+720q^7-3024q^8\\
&q^{\frac{1}{3}}\left(1+4q-40q^2-36q^3+638q^4-1136q^5-556q^6+3384q^7-3089q^8\right)\\
&-6q^{\frac{2}{3}}\left(1-q-28q^2+94q^3-40q^4-147q^5-4q^6+140q^7+638q^8\right)\\
\end{aligned}
\end{pmatrix}\\
Y_{\mathbf{3}_2}^{(6)}(\tau)
&=
\begin{pmatrix}
\begin{aligned}
&6q^{\frac{1}{2}}\left(1+69q+54q^2-88q^3-1071q^4+540q^5-418q^6+3726q^7+594q^8\right)\\
&54q^{\frac{5}{6}}\left(1+10q+11q^2-76q^3-11q^4-88q^5+319q^6-24q^7+361q^8\right)\\
&q^{\frac{1}{6}}\left(1-88q-418q^2+836q^3-209q^4+4256q^5-298q^6-12100q^7-9063q^8\right)\\
\end{aligned}
\end{pmatrix}\\
\end{align*}

\begin{align*}
\text{Weight 7:}\\
Y_{\mathbf{1}_1}^{(7)}(\tau)
&=q^{\frac{1}{4}}\left(1+234q+729q^2-4070q^3-990q^4+39131q^6-31878q^7\right)\\
Y_{\mathbf{1}_1'}^{(7)}(\tau)
&=q^{\frac{7}{12}}\left(1-14q+77q^2-182q^3+924q^5-1547q^6-506q^7+3003q^8\right)\\
Y_{\mathbf{1}_3'}^{(7)}(\tau)
&=q^{\frac{1}{12}}\left(1-506q-15625q^2-89206q^3-271151q^4-420838q^5-638066q^6+56446q^8\right)\\
Y_{\mathbf{3}_1}^{(7)}(\tau)
&=
\begin{pmatrix}
\begin{aligned}
&q^{\frac{1}{4}}\left(1-90q-243q^2+1762q^3-1638q^4+2916q^5-7525q^6-16002q^7+26244q^8\right)\\
&6q^{\frac{7}{12}}\left(1+67q-85q^2-263q^3-810q^4+2949q^5-413q^6+1762q^7-6717q^8\right)\\
&54q^{\frac{11}{12}}\left(1+8q-10q^2-106q^3+151q^4+44q^5+383q^6-670q^7-273q^8\right)\\
\end{aligned}
\end{pmatrix}\\
Y_{\mathbf{3}_3^{I}}^{(7)}(\tau)
&=
\begin{pmatrix}
\begin{aligned}
&-162q^{\frac{3}{4}}\left(1+24q+216q^2+882q^3+1608q^4+1728q^5+2673q^6-2040q^7\right.\\
&\qquad\qquad\qquad\qquad\qquad\qquad\qquad\qquad\qquad\qquad\qquad\qquad\qquad\qquad\quad\left.-2160q^8\right)\\
&q^{\frac{1}{12}}\left(1-182q+8027q^2+22250q^3+94321q^4+149402q^5+209842q^6\right.\\
&\qquad\qquad\qquad\qquad\qquad\qquad\qquad\qquad\qquad\qquad\qquad\quad\left.+21060q^7-119486q^8\right)\\
&18q^{\frac{5}{12}}\left(1-79q-1183q^2-5223q^3-10743q^4-23510q^5-11664q^6\right.\\
&\qquad\qquad\qquad\qquad\qquad\qquad\qquad\qquad\qquad\qquad\qquad\qquad\left.-4799q^7+36537q^8\right)\\
\end{aligned}
\end{pmatrix}\\
Y_{\mathbf{3}_3^{II}}^{(7)}(\tau)
&=
\begin{pmatrix}
\begin{aligned}
&-18q^{\frac{3}{4}}\left(1+240q+2160q^2+6714q^3+16080q^4+17280q^5+20169q^6\right.\\
&\qquad\qquad\qquad\qquad\qquad\qquad\qquad\qquad\qquad\qquad\qquad\qquad\left.-20400q^7-21600q^8\right)\\
&q^{\frac{1}{12}}\left(1+250q+4571q^2+30890q^3+91729q^4+139898q^5+185650q^6\right.\\
&\qquad\qquad\qquad\qquad\qquad\qquad\qquad\qquad\qquad\qquad\qquad\qquad\left.+49140q^7+30850q^8\right)\\
&-6q^{\frac{5}{12}}\left(1+245q+3353q^2+15837q^3+35265q^4+60730q^5+38880q^6\right.\\
&\qquad\qquad\qquad\qquad\qquad\qquad\qquad\qquad\qquad\qquad\qquad\qquad\left.+41533q^7-153327q^8\right)
\end{aligned}
\end{pmatrix}
\end{align*}

\begin{align*}
\text{Weight 8 :}\\
Y_{\mathbf{1}_0}^{(8)}(\tau)
&=1+480q+61920q^2+1050240q^3+7926240q^4+37500480q^5+135480960q^6\\
&\qquad\qquad\qquad\qquad\qquad\qquad\qquad\qquad\qquad\qquad\qquad+395301120q^7+1014559200q^8\\
Y_{\mathbf{1}_0'}^{(8)}(\tau)
&=q^{\frac{1}{3}}\left(1+232q+260q^2-5760q^3+6890q^4+7744q^5+33176q^6\right.\\
&\qquad\qquad\qquad\qquad\qquad\qquad\qquad\qquad\qquad\qquad\qquad\qquad\qquad\left.-115200q^7+14035q^8\right)\\
Y_{\mathbf{1}_0''}^{(8)}(\tau)
&=q^{\frac{2}{3}}\left(1-16q+104q^2-320q^3+260q^4+1248q^5-3712q^6+1664q^7+6890q^8\right)\\
Y_{\mathbf{1}_2''}^{(8)}(\tau)
&=q^{\frac{1}{6}}\left(1-508q-14614q^2-57448q^3-78125q^4+178916q^5+279710q^6\right.\\
&\qquad\qquad\qquad\qquad\qquad\qquad\qquad\qquad\qquad\qquad\qquad\qquad\quad\left.+1035224q^7-565479q^8\right)
\end{align*}
\begin{align*}
Y_{\mathbf{3}_0^{I}}^{(8)}(\tau)
&=
\begin{pmatrix}
\begin{aligned}
&1-6q-19728q^2-353814q^3-2657868q^4-12447684q^5-45130248q^6\\
&\qquad\qquad\qquad\qquad\qquad\qquad\qquad\qquad\qquad\qquad\quad-131857248q^7-338596020q^8\\
&3q^{\frac{1}{3}}\left(1+718q+40112q^2+493848q^3+3056054q^4+13201186q^5+43595300q^6\right.\\
&\qquad\qquad\qquad\qquad\qquad\qquad\qquad\qquad\qquad\qquad\quad\left.+122621184q^7+297806647q^8\right)\\
&18q^{\frac{2}{3}}\left(2+697q+16975q^2+158282q^3+863656q^4+3340587q^5+10482886q^6\right.\\
&\qquad\qquad\qquad\qquad\qquad\qquad\qquad\qquad\qquad\qquad\qquad\left.+27679084q^7+65804572q^8\right)\\
\end{aligned}
\end{pmatrix}\\
Y_{\mathbf{3}_0^{II}}^{(8)}(\tau)
&=
\begin{pmatrix}
\begin{aligned}
&54q+324q^2-1458q^3-4968q^4+21060q^5-8748q^6-3456q^7-71280q^8\\
&q^{\frac{1}{3}}\left(1-92q-64q^2+2340q^3-5098q^4+3856q^5-8620q^6-5688q^7+73975q^8\right)\\
&6q^{\frac{2}{3}}\left(1+65q-220q^2-158q^3-64q^4+4731q^5-5980q^6-2548q^7-5098q^8\right)\\
\end{aligned}
\end{pmatrix}\\
Y_{\mathbf{3}_2^{I}}^{(8)}(\tau)
&=
\begin{pmatrix}
\begin{aligned}
&-18q^{\frac{1}{2}}\left(1+81q+918q^2+3056q^3+729q^4-8964q^5-24010q^6\right.\\
&\qquad\qquad\qquad\qquad\qquad\qquad\qquad\qquad\qquad\qquad\qquad\qquad\quad\left.+4374q^7-50814q^8\right)\\
&54q^{\frac{5}{6}}\left(1-86q-589q^2-436q^3-971q^4+5648q^5+6655q^6+10704q^7-20927q^8\right)\\
&q^{\frac{1}{6}}\left(1+464q+1910q^2+29060q^3+18103q^4-74776q^5-112978q^6\right.\\
&\qquad\qquad\qquad\qquad\qquad\qquad\qquad\qquad\qquad\qquad\qquad\qquad\left.-198244q^7+231561q^8\right)\\
\end{aligned}
\end{pmatrix}\\
Y_{\mathbf{3}_2^{II}}^{(8)}(\tau)
&=
\begin{pmatrix}
\begin{aligned}
&6q^{\frac{1}{2}}\left(1-243q-2970q^2-8608q^3+729q^4+10476q^5+92630q^6\right.\\
&\qquad\qquad\qquad\qquad\qquad\qquad\qquad\qquad\qquad\qquad\qquad\qquad\quad\left.+21870q^7+34722q^8\right)\\
&-18q^{\frac{5}{6}}\left(5+218q+1591q^2+3004q^3-967q^4-17120q^5-8845q^6\right.\\
&\qquad\qquad\qquad\qquad\qquad\qquad\qquad\qquad\qquad\qquad\qquad\qquad\quad\left.-38496q^7+54773q^8\right)\\
&q^{\frac{1}{6}}\left(1+32q+6230q^2+13940q^3+33655q^4-55336q^5-125074q^6\right.\\
&\qquad\qquad\qquad\qquad\qquad\qquad\qquad\qquad\qquad\qquad\qquad\qquad\left.-279892q^7+110601q^8\right)
\end{aligned}
\end{pmatrix}
\end{align*}
\begin{align*}
\text{Weight 9:}\\
Y_{\mathbf{1}_1}^{(9)}(\tau)
&=q^{\frac{1}{4}}\left(1-510q-13599q^2-27710q^3+50370q^4+362880q^5-130525q^6\right.\\
&\left.\qquad\qquad\qquad\qquad\qquad\qquad\qquad\qquad\qquad\qquad\qquad\qquad\qquad\qquad+54978q^7-2721600q^8\right)\\
Y_{\mathbf{1}_3}^{(9)}(\tau)
&=q^{\frac{3}{4}}\left(1-18q+135q^2-510q^3+765q^4+1242q^5-7038q^6+8280q^7+9180q^8\right)\\
Y_{\mathbf{1}_3'}^{(9)}(\tau)
&=q^{\frac{1}{12}}\left(1+478q+60959q^2+925922q^3+5764801q^4+20722082q^5+54717118q^6\right.\\
&\qquad\qquad\qquad\qquad\qquad\qquad\qquad\qquad\qquad\qquad\qquad\qquad\left.+103864320q^7+173379838q^8\right)\\
Y_{\mathbf{1}_3''}^{(9)}(\tau)
&=q^{\frac{5}{12}}\left(1+230q-205q^2-6510q^3+18615q^4+478q^5-181220q^7+221595q^8\right)\\
Y_{\mathbf{3}_1^{I}}^{(9)}(\tau)
&=
\begin{pmatrix}
\begin{aligned}
&q^{\frac{1}{4}}\left(1+138q+4869q^2+7282q^3-10542q^4-127008q^5+9443q^6\right.\\
&\qquad\qquad\qquad\qquad\qquad\qquad\qquad\qquad\qquad\qquad\qquad\qquad\left.+132090q^7+716364q^8\right)\\
&-1458q^{\frac{7}{12}}\left(q+10q^2+11q^3-82q^4-71q^5-154q^6+784q^7+132q^8\right)\\
&-54q^{\frac{11}{12}}\left(1+74q+392q^2-304q^3-2111q^4-5386q^5+9923q^6\right.\\
&\qquad\qquad\qquad\qquad\qquad\qquad\qquad\qquad\qquad\qquad\qquad\qquad\qquad\left.+2942q^7+36645q^8\right)\\
\end{aligned}
\end{pmatrix}
\end{align*}
\begin{align*}
Y_{\mathbf{3}_1^{II}}^{(9)}(\tau)
&=
\begin{pmatrix}
\begin{aligned}
&-18q^{\frac{1}{4}}\left(q-12q^2+54q^3-94q^4-27q^5+216q^6+119q^7-162q^8\right)\\
&q^{\frac{7}{12}}\left(1-2q-55q^2+250q^3+504q^4-5808q^5+12853q^6+4042q^7-59169q^8\right)\\
&-6q^{\frac{11}{12}}\left(1-7q-13q^2+263q^3-896q^4+689q^5+2309q^6-4348q^7-858q^8\right)\\
\end{aligned}
\end{pmatrix}\\
Y_{\mathbf{3}_3^{I}}^{(9)}(\tau)
&=
\begin{pmatrix}
\begin{aligned}
&-18q^{\frac{3}{4}}\left(5-252q-18036q^2-122106q^3-527292q^4-1457136q^5-3383487q^6\right.\\
&\qquad\qquad\qquad\qquad\qquad\qquad\qquad\qquad\qquad\qquad\qquad\qquad\qquad\left.-5264100q^7-7682472q^8\right)\\
&q^{\frac{1}{12}}\left(1-1142q-24253q^2-268990q^3-1882895q^4-7513870q^5-17288318q^6\right.\\
&\qquad\qquad\qquad\qquad\qquad\qquad\qquad\qquad\qquad\qquad\qquad\qquad\quad\left.-34482708q^7-57510014q^8\right)\\
&-18q^{\frac{5}{12}}\left(1-229q-6307q^2-65829q^3-349827q^4-1099286q^5-2535624q^6\right.\\
&\qquad\qquad\qquad\qquad\qquad\qquad\qquad\qquad\qquad\qquad\qquad\qquad\qquad\left.-4819145q^7-6437523q^8\right)\\
\end{aligned}
\end{pmatrix}\\
Y_{\mathbf{3}_3^{II}}^{(9)}(\tau)
&=
\begin{pmatrix}
\begin{aligned}
&54q^{\frac{3}{4}}\left(1+252q+5076q^2+41286q^3+177372q^4+498096q^5+1085733q^6\right.\\
&\qquad\qquad\qquad\qquad\qquad\qquad\qquad\qquad\qquad\qquad\qquad\qquad\qquad\left.+1764900q^7+2602152q^8\right)\\
&q^{\frac{1}{12}}\left(1+154q-19069q^2-320830q^3-1937327q^4-6718126q^5-18449534q^6\right.\\
&\qquad\qquad\qquad\qquad\qquad\qquad\qquad\qquad\qquad\qquad\qquad\qquad\quad\left.-34911684q^7-58038782q^8\right)\\
&6q^{\frac{5}{12}}\left(1+311q+19397q^2+207735q^3+1009893q^4+3343834q^5+7579656q^6\right.\\
&\qquad\qquad\qquad\qquad\qquad\qquad\qquad\qquad\qquad\qquad\qquad\qquad\quad\left.+14457019q^7+19718133q^8\right)\\
\end{aligned}
\end{pmatrix}
\end{align*}

\bibliographystyle{JHEP-related/JHEP}
\bibliography{biblio.bib}

@inbook{Feruglio:2017spp,
    author = "Feruglio, Ferruccio",
    editor = "Levy, Aharon and Forte, Stefano and Ridolfi, Giovanni",
    title = "{Are neutrino masses modular forms?}",
    booktitle = "{From My Vast Repertoire ...}: {Guido Altarelli's Legacy}",
    eprint = "1706.08749",
    archivePrefix = "arXiv",
    primaryClass = "hep-ph",
    reportNumber = "DFPD-2017-TH-09",
    doi = "10.1142/9789813238053_0012",
    pages = "227--266",
    year = "2019"
}

@article{Ferrara:1989bc,
    author = "Ferrara, S. and Lust, D. and Shapere, Alfred D. and Theisen, S.",
    title = "{Modular Invariance in Supersymmetric Field Theories}",
    reportNumber = "IASSNS-HEP-89/25, CERN-TH-5363/89, UCLA/89/TEP/19",
    doi = "10.1016/0370-2693(89)90583-2",
    journal = "Phys. Lett. B",
    volume = "225",
    pages = "363",
    year = "1989"
}

@article{Ferrara:1989qb,
    author = "Ferrara, S. and Lust, . D. and Theisen, S.",
    title = "{Target Space Modular Invariance and Low-Energy Couplings in Orbifold Compactifications}",
    reportNumber = "CERN-TH-5545-89, UCLA-89-TEP-42",
    doi = "10.1016/0370-2693(89)90631-X",
    journal = "Phys. Lett. B",
    volume = "233",
    pages = "147--152",
    year = "1989"
}

@article{Giveon:1994fu,
    author = "Giveon, Amit and Porrati, Massimo and Rabinovici, Eliezer",
    title = "{Target space duality in string theory}",
    eprint = "hep-th/9401139",
    archivePrefix = "arXiv",
    reportNumber = "RI-1-94, NYU-TH-94-01-01",
    doi = "10.1016/0370-1573(94)90070-1",
    journal = "Phys. Rept.",
    volume = "244",
    pages = "77--202",
    year = "1994"
}

@article{Dixon:1986qv,
    author = "Dixon, Lance J. and Friedan, Daniel and Martinec, Emil J. and Shenker, Stephen H.",
    title = "{The Conformal Field Theory of Orbifolds}",
    reportNumber = "EFI-86-42-CHICAGO",
    doi = "10.1016/0550-3213(87)90676-6",
    journal = "Nucl. Phys. B",
    volume = "282",
    pages = "13--73",
    year = "1987"
}

@article{Erler:1991nr,
    author = "Erler, J. and Jungnickel, D. and Lauer, J.",
    title = "{Dependence of Yukawa couplings on the axionic background moduli of Z(N) orbifolds}",
    reportNumber = "SLAC-PUB-5613, MPI-PH-91-65, TUM-TH-123-91",
    doi = "10.1103/PhysRevD.45.3651",
    journal = "Phys. Rev. D",
    volume = "45",
    pages = "3651--3668",
    year = "1992"
}

@article{Dixon:1989fj,
    author = "Dixon, Lance J. and Kaplunovsky, Vadim and Louis, Jan",
    title = "{On Effective Field Theories Describing (2,2) Vacua of the Heterotic String}",
    reportNumber = "SLAC-PUB-4959, UTTG-19-89",
    doi = "10.1016/0550-3213(90)90057-K",
    journal = "Nucl. Phys. B",
    volume = "329",
    pages = "27--82",
    year = "1990"
}

@article{deAdelhartToorop:2011re,
    author = "de Adelhart Toorop, Reinier and Feruglio, Ferruccio and Hagedorn, Claudia",
    title = "{Finite Modular Groups and Lepton Mixing}",
    eprint = "1112.1340",
    archivePrefix = "arXiv",
    primaryClass = "hep-ph",
    reportNumber = "DFPD-11-TH-18",
    doi = "10.1016/j.nuclphysb.2012.01.017",
    journal = "Nucl. Phys. B",
    volume = "858",
    pages = "437--467",
    year = "2012"
}

@article{Feruglio:2019ybq,
    author = "Feruglio, Ferruccio and Romanino, Andrea",
    title = "{Lepton flavor symmetries}",
    eprint = "1912.06028",
    archivePrefix = "arXiv",
    primaryClass = "hep-ph",
    doi = "10.1103/RevModPhys.93.015007",
    journal = "Rev. Mod. Phys.",
    volume = "93",
    number = "1",
    pages = "015007",
    year = "2021"
}

@article{Ding:2023htn,
    author = "Ding, Gui-Jun and King, Stephen F.",
    title = "{Neutrino mass and mixing with modular symmetry}",
    eprint = "2311.09282",
    archivePrefix = "arXiv",
    primaryClass = "hep-ph",
    doi = "10.1088/1361-6633/ad52a3",
    journal = "Rept. Prog. Phys.",
    volume = "87",
    number = "8",
    pages = "084201",
    year = "2024"
}

@article{Feruglio:2025ztj,
    author = "Feruglio, Ferruccio and Ramos-Sanchez, Saul",
    editor = "Fischer, Christian S. and Guo, Feng-Kun and Pich, Antonio",
    title = "{Quark and lepton masses}",
    eprint = "2506.20755",
    archivePrefix = "arXiv",
    primaryClass = "hep-ph",
    doi = "10.1016/B978-0-443-26598-3.00031-6",
    volume = "2",
    pages = "542--571",
    year = "2026"
}

@article{Binetruy:1995nt,
    author = "Binetruy, Pierre and Dudas, Emilian",
    title = "{Dynamical mass matrices from effective superstring theories}",
    eprint = "hep-ph/9505295",
    archivePrefix = "arXiv",
    reportNumber = "LPTHE-ORSAY-95-18, SACLAY-SPH-T-95-042, SPHT-SACLAY-T95-042",
    doi = "10.1016/0550-3213(95)00345-S",
    journal = "Nucl. Phys. B",
    volume = "451",
    pages = "31--52",
    year = "1995"
}

@article{Dudas:1995eq,
    author = "Dudas, E. and Pokorski, S. and Savoy, Carlos A.",
    title = "{Soft scalar masses in supergravity with horizontal U(1)-x gauge symmetry}",
    eprint = "hep-ph/9509410",
    archivePrefix = "arXiv",
    reportNumber = "SACLAY-DPHT-95-094, MPI-PHT-95-89",
    doi = "10.1016/0370-2693(95)01536-1",
    journal = "Phys. Lett. B",
    volume = "369",
    pages = "255--261",
    year = "1996"
}

@article{Dudas:1996fe,
    author = "Dudas, E. and Grojean, C. and Pokorski, S. and Savoy, Carlos A.",
    title = "{Abelian flavor symmetries in supersymmetric models}",
    eprint = "hep-ph/9606383",
    archivePrefix = "arXiv",
    reportNumber = "SACLAY-SPH-T-96-065",
    doi = "10.1016/S0550-3213(96)90123-6",
    journal = "Nucl. Phys. B",
    volume = "481",
    pages = "85--108",
    year = "1996"
}

@article{Leontaris:1997vw,
    author = "Leontaris, G. K. and Tracas, N. D.",
    title = "{Modular weights, U(1)'s and mass matrices}",
    eprint = "hep-ph/9709510",
    archivePrefix = "arXiv",
    reportNumber = "NTUA-65-97A, IOA-TH-97-005",
    doi = "10.1016/S0370-2693(97)01412-3",
    journal = "Phys. Lett. B",
    volume = "419",
    pages = "206--210",
    year = "1998"
}

@article{Kobayashi:2018wkl,
    author = "Kobayashi, Tatsuo and Shimizu, Yusuke and Takagi, Kenta and Tanimoto, Morimitsu and Tatsuishi, Takuya H. and Uchida, Hikaru",
    title = "{Finite modular subgroups for fermion mass matrices and baryon/lepton number violation}",
    eprint = "1812.11072",
    archivePrefix = "arXiv",
    primaryClass = "hep-ph",
    reportNumber = "EPHOU-18-017, HUPD1810",
    doi = "10.1016/j.physletb.2019.05.034",
    journal = "Phys. Lett. B",
    volume = "794",
    pages = "114--121",
    year = "2019"
}

@article{Penedo:2018nmg,
    author = "Penedo, J. T. and Petcov, S. T.",
    title = "{Lepton Masses and Mixing from Modular $S_4$ Symmetry}",
    eprint = "1806.11040",
    archivePrefix = "arXiv",
    primaryClass = "hep-ph",
    reportNumber = "SISSA 25/2018/FISI, IPMU18-0121, SISSA-25-2018-FISI",
    doi = "10.1016/j.nuclphysb.2018.12.016",
    journal = "Nucl. Phys. B",
    volume = "939",
    pages = "292--307",
    year = "2019"
}

@article{Criado:2018thu,
    author = "Criado, Juan Carlos and Feruglio, Ferruccio",
    title = "{Modular Invariance Faces Precision Neutrino Data}",
    eprint = "1807.01125",
    archivePrefix = "arXiv",
    primaryClass = "hep-ph",
    doi = "10.21468/SciPostPhys.5.5.042",
    journal = "SciPost Phys.",
    volume = "5",
    number = "5",
    pages = "042",
    year = "2018"
}

@article{Ding:2019zxk,
    author = "Ding, Gui-Jun and King, Stephen F. and Liu, Xiang-Gan",
    title = "{Modular A$_{4}$ symmetry models of neutrinos and charged leptons}",
    eprint = "1907.11714",
    archivePrefix = "arXiv",
    primaryClass = "hep-ph",
    reportNumber = "USTC-ICTS-19-16",
    doi = "10.1007/JHEP09(2019)074",
    journal = "JHEP",
    volume = "09",
    pages = "074",
    year = "2019"
}

@article{Yao:2020qyy,
    author = "Yao, Chang-Yuan and Lu, Jun-Nan and Ding, Gui-Jun",
    title = "{Modular Invariant $A_{4}$ Models for Quarks and Leptons with Generalized CP Symmetry}",
    eprint = "2012.13390",
    archivePrefix = "arXiv",
    primaryClass = "hep-ph",
    reportNumber = "USTC-ICTS/PCFT-20-21",
    doi = "10.1007/JHEP05(2021)102",
    journal = "JHEP",
    volume = "05",
    pages = "102",
    year = "2021"
}

@article{Okada:2020brs,
    author = "Okada, Hiroshi and Tanimoto, Morimitsu",
    title = "{Spontaneous CP violation by modulus $\tau$ in $A_4$ model of lepton flavors}",
    eprint = "2012.01688",
    archivePrefix = "arXiv",
    primaryClass = "hep-ph",
    reportNumber = "APCTP Pre2020-033",
    doi = "10.1007/JHEP03(2021)010",
    journal = "JHEP",
    volume = "03",
    pages = "010",
    year = "2021"
}

@article{Novichkov:2018nkm,
    author = "Novichkov, P. P. and Penedo, J. T. and Petcov, S. T. and Titov, A. V.",
    title = "{Modular A$_{5}$ symmetry for flavour model building}",
    eprint = "1812.02158",
    archivePrefix = "arXiv",
    primaryClass = "hep-ph",
    reportNumber = "SISSA 54/2018/FISI, IPMU18-0202, IPPP/18/105",
    doi = "10.1007/JHEP04(2019)174",
    journal = "JHEP",
    volume = "04",
    pages = "174",
    year = "2019"
}

@article{Liu:2019khw,
    author = "Liu, Xiang-Gan and Ding, Gui-Jun",
    title = "{Neutrino Masses and Mixing from Double Covering of Finite Modular Groups}",
    eprint = "1907.01488",
    archivePrefix = "arXiv",
    primaryClass = "hep-ph",
    reportNumber = "USTC-ICTS-19-17",
    doi = "10.1007/JHEP08(2019)134",
    journal = "JHEP",
    volume = "08",
    pages = "134",
    year = "2019"
}

@article{Liu:2020msy,
    author = "Liu, Xiang-Gan and Yao, Chang-Yuan and Qu, Bu-Yao and Ding, Gui-Jun",
    title = "{Half-integral weight modular forms and application to neutrino mass models}",
    eprint = "2007.13706",
    archivePrefix = "arXiv",
    primaryClass = "hep-ph",
    reportNumber = "USTC-ICTS/PCFT-20-22",
    doi = "10.1103/PhysRevD.102.115035",
    journal = "Phys. Rev. D",
    volume = "102",
    number = "11",
    pages = "115035",
    year = "2020"
}

@article{Yao:2020zml,
    author = "Yao, Chang-Yuan and Liu, Xiang-Gan and Ding, Gui-Jun",
    title = "{Fermion masses and mixing from the double cover and metaplectic cover of the $A_5$ modular group}",
    eprint = "2011.03501",
    archivePrefix = "arXiv",
    primaryClass = "hep-ph",
    reportNumber = "USTC-ICTS/PCFT-20-36",
    doi = "10.1103/PhysRevD.103.095013",
    journal = "Phys. Rev. D",
    volume = "103",
    number = "9",
    pages = "095013",
    year = "2021"
}

@article{Novichkov:2018ovf,
    author = "Novichkov, P. P. and Penedo, J. T. and Petcov, S. T. and Titov, A. V.",
    title = "{Modular S$_{4}$ models of lepton masses and mixing}",
    eprint = "1811.04933",
    archivePrefix = "arXiv",
    primaryClass = "hep-ph",
    reportNumber = "SISSA 47/2018/FISI, IPMU18-0187, IPPP/18/98",
    doi = "10.1007/JHEP04(2019)005",
    journal = "JHEP",
    volume = "04",
    pages = "005",
    year = "2019"
}

@article{Novichkov:2019sqv,
    author = "Novichkov, P. P. and Penedo, J. T. and Petcov, S. T. and Titov, A. V.",
    title = "{Generalised CP Symmetry in Modular-Invariant Models of Flavour}",
    eprint = "1905.11970",
    archivePrefix = "arXiv",
    primaryClass = "hep-ph",
    reportNumber = "SISSA 14/2019/FISI, IPMU19-0080, IPPP/19/46, CFTP/19-019",
    doi = "10.1007/JHEP07(2019)165",
    journal = "JHEP",
    volume = "07",
    pages = "165",
    year = "2019"
}

@article{Novichkov:2020eep,
    author = "Novichkov, P. P. and Penedo, J. T. and Petcov, S. T.",
    title = "{Double cover of modular $S_4$ for flavour model building}",
    eprint = "2006.03058",
    archivePrefix = "arXiv",
    primaryClass = "hep-ph",
    reportNumber = "SISSA 12/2020/FISI, IPMU20-0065, CFTP/20-006",
    doi = "10.1016/j.nuclphysb.2020.115301",
    journal = "Nucl. Phys. B",
    volume = "963",
    pages = "115301",
    year = "2021"
}

@article{Liu:2020akv,
    author = "Liu, Xiang-Gan and Yao, Chang-Yuan and Ding, Gui-Jun",
    title = "{Modular invariant quark and lepton models in double covering of $S_4$ modular group}",
    eprint = "2006.10722",
    archivePrefix = "arXiv",
    primaryClass = "hep-ph",
    reportNumber = "USTC-ICTS/PCFT-20-18",
    doi = "10.1103/PhysRevD.103.056013",
    journal = "Phys. Rev. D",
    volume = "103",
    number = "5",
    pages = "056013",
    year = "2021"
}

@article{Ding:2019xna,
    author = "Ding, Gui-Jun and King, Stephen F. and Liu, Xiang-Gan",
    title = "{Neutrino mass and mixing with $A_5$ modular symmetry}",
    eprint = "1903.12588",
    archivePrefix = "arXiv",
    primaryClass = "hep-ph",
    reportNumber = "USTC-ICTS-19-03",
    doi = "10.1103/PhysRevD.100.115005",
    journal = "Phys. Rev. D",
    volume = "100",
    number = "11",
    pages = "115005",
    year = "2019"
}

@article{Wang:2020lxk,
    author = "Wang, Xin and Yu, Bingrong and Zhou, Shun",
    title = "{Double covering of the modular $A_5$ group and lepton flavor mixing in the minimal seesaw model}",
    eprint = "2010.10159",
    archivePrefix = "arXiv",
    primaryClass = "hep-ph",
    doi = "10.1103/PhysRevD.103.076005",
    journal = "Phys. Rev. D",
    volume = "103",
    number = "7",
    pages = "076005",
    year = "2021"
}

@article{Wang:2021mkw,
    author = "Wang, Xin and Zhou, Shun",
    title = "{Explicit perturbations to the stabilizer {\ensuremath{\tau}} = i of modular $ {A}_5^{\prime } $ symmetry and leptonic CP violation}",
    eprint = "2102.04358",
    archivePrefix = "arXiv",
    primaryClass = "hep-ph",
    doi = "10.1007/JHEP07(2021)093",
    journal = "JHEP",
    volume = "07",
    pages = "093",
    year = "2021"
}

@article{Li:2021buv,
    author = "Li, Cai-Chang and Liu, Xiang-Gan and Ding, Gui-Jun",
    title = "{Modular symmetry at level 6 and a new route towards finite modular groups}",
    eprint = "2108.02181",
    archivePrefix = "arXiv",
    primaryClass = "hep-ph",
    reportNumber = "USTC-ICTS/PCFT-21-27",
    doi = "10.1007/JHEP10(2021)238",
    journal = "JHEP",
    volume = "10",
    pages = "238",
    year = "2021"
}

@article{Ding:2020msi,
    author = "Ding, Gui-Jun and King, Stephen F. and Li, Cai-Chang and Zhou, Ye-Ling",
    title = "{Modular Invariant Models of Leptons at Level 7}",
    eprint = "2004.12662",
    archivePrefix = "arXiv",
    primaryClass = "hep-ph",
    reportNumber = "USTC-ICTS/PCFT-20-11",
    doi = "10.1007/JHEP08(2020)164",
    journal = "JHEP",
    volume = "08",
    pages = "164",
    year = "2020"
}

@article{Feruglio:2021dte,
    author = "Feruglio, Ferruccio and Gherardi, Valerio and Romanino, Andrea and Titov, Arsenii",
    title = "{Modular invariant dynamics and fermion mass hierarchies around $\tau = i$}",
    eprint = "2101.08718",
    archivePrefix = "arXiv",
    primaryClass = "hep-ph",
    reportNumber = "SISSA 05/2021/FISI, FTUV-21-0119.3119, IFIC/21-02",
    doi = "10.1007/JHEP05(2021)242",
    journal = "JHEP",
    volume = "05",
    pages = "242",
    year = "2021"
}

@article{Feruglio:2022koo,
    author = "Feruglio, Ferruccio",
    title = "{Universal Predictions of Modular Invariant Flavor Models near the Self-Dual Point}",
    eprint = "2211.00659",
    archivePrefix = "arXiv",
    primaryClass = "hep-ph",
    doi = "10.1103/PhysRevLett.130.101801",
    journal = "Phys. Rev. Lett.",
    volume = "130",
    number = "10",
    pages = "101801",
    year = "2023"
}

@article{Feruglio:2023mii,
    author = "Feruglio, Ferruccio",
    title = "{Fermion masses, critical behavior and universality}",
    eprint = "2302.11580",
    archivePrefix = "arXiv",
    primaryClass = "hep-ph",
    doi = "10.1007/JHEP03(2023)236",
    journal = "JHEP",
    volume = "03",
    pages = "236",
    year = "2023"
}

@article{King:2020qaj,
    author = "King, Simon J. D. and King, Stephen F.",
    title = "{Fermion mass hierarchies from modular symmetry}",
    eprint = "2002.00969",
    archivePrefix = "arXiv",
    primaryClass = "hep-ph",
    doi = "10.1007/JHEP09(2020)043",
    journal = "JHEP",
    volume = "09",
    pages = "043",
    year = "2020"
}

@article{Okada:2020ukr,
    author = "Okada, Hiroshi and Tanimoto, Morimitsu",
    title = "{Modular invariant flavor model of $A_4$ and hierarchical structures at nearby fixed points}",
    eprint = "2009.14242",
    archivePrefix = "arXiv",
    primaryClass = "hep-ph",
    reportNumber = "APCTP Pre2020-024",
    doi = "10.1103/PhysRevD.103.015005",
    journal = "Phys. Rev. D",
    volume = "103",
    number = "1",
    pages = "015005",
    year = "2021"
}

@article{Novichkov:2021evw,
    author = "Novichkov, P. P. and Penedo, J. T. and Petcov, S. T.",
    title = "{Fermion mass hierarchies, large lepton mixing and residual modular symmetries}",
    eprint = "2102.07488",
    archivePrefix = "arXiv",
    primaryClass = "hep-ph",
    reportNumber = "SISSA 09/2021/FISI, IPMU21-0009, CFTP/21-003",
    doi = "10.1007/JHEP04(2021)206",
    journal = "JHEP",
    volume = "04",
    pages = "206",
    year = "2021"
}

@article{Liu:2021gwa,
    author = "Liu, Xiang-Gan and Ding, Gui-Jun",
    title = "{Modular flavor symmetry and vector-valued modular forms}",
    eprint = "2112.14761",
    archivePrefix = "arXiv",
    primaryClass = "hep-ph",
    reportNumber = "USTC-ICTS/PCFT-21-41",
    doi = "10.1007/JHEP03(2022)123",
    journal = "JHEP",
    volume = "03",
    pages = "123",
    year = "2022"
}

@article{Ding:2023ydy,
    author = "Ding, Gui-Jun and Liu, Xiang-Gan and Lu, Jun-Nan and Weng, Ming-Hua",
    title = "{Modular binary octahedral symmetry for flavor structure of Standard Model}",
    eprint = "2307.14926",
    archivePrefix = "arXiv",
    primaryClass = "hep-ph",
    reportNumber = "UCI-TR-2023-06",
    doi = "10.1007/JHEP11(2023)083",
    journal = "JHEP",
    volume = "11",
    pages = "083",
    year = "2023"
}

@article{Arriaga-Osante:2023wnu,
    author = "Arriaga-Osante, Carlos and Liu, Xiang-Gan and Ramos-Sanchez, Saul",
    title = "{Quark and lepton modular models from the binary dihedral flavor symmetry}",
    eprint = "2311.10136",
    archivePrefix = "arXiv",
    primaryClass = "hep-ph",
    doi = "10.1007/JHEP05(2024)119",
    journal = "JHEP",
    volume = "05",
    pages = "119",
    year = "2024"
}

@incollection{ConderDobcsanyi2006,
  author    = {Conder, Marston and Dobcs{\'a}nyi, Peter},
  title     = {Normal subgroups of low index in the modular group and other {Hecke} groups},
  booktitle = {Combinatorial Group Theory, Number Theory and Discrete Groups},
  series    = {Contemporary Mathematics},
  volume    = {421},
  pages     = {91--114},
  year      = {2006},
  publisher = {American Mathematical Society},
  address   = {Providence, RI}
}

@article{CentellesChulia:2026bkr,
    author = "Centelles Chuli{\'a}, Salvador and Li, Xueqi and Liu, Xiang-Gan and Medina, Omar and Penedo, J. T.",
    title = "{Fermion mass relations in one-parameter modular models}",
    eprint = "2606.25027",
    archivePrefix = "arXiv",
    primaryClass = "hep-ph",
    month = "6",
    year = "2026"
}

@misc{candelori2016vectorvaluedmodularforms,
      title={Vector valued modular forms and the modular orbifold of elliptic curves}, 
      author={Luca Candelori and Cameron Franc},
      year={2016},
      eprint={1506.09192},
      archivePrefix={arXiv},
      primaryClass={math.NT},
      url={https://arxiv.org/abs/1506.09192}, 
}

@book{cohen2017modular,
  title={Modular Forms: A Classical Approach},
  author={Cohen, Henri and Str{\"o}mberg, Fredrik},
  volume={179},
  year={2017},
  publisher={American Mathematical Society},
  series={Graduate Studies in Mathematics},
  address={Providence, Rhode Island},
  isbn={978-0-8218-4947-7}
}

@book{bruinier2008one,
  author    = {Bruinier, Jan Hendrik and van der Geer, Gerard and Harder, G{\"u}nter and Zagier, Don},
  editor    = {Ranestad, Kristian},
  title     = {The 1-2-3 of Modular Forms: Lectures at a Summer School in Nordfjordeid, Norway},
  series    = {Universitext},
  publisher = {Springer Berlin Heidelberg},
  year      = {2008},
  isbn      = {978-3-540-74117-6},
  doi       = {10.1007/978-3-540-74119-0}
}

@article{zbMATH06664578,
 author = {Franc, Cameron and Mason, Geoffrey},
 title = {Hypergeometric series, modular linear differential equations and vector-valued modular forms},
 fjournal = {The Ramanujan Journal},
 journal = {Ramanujan J.},
 issn = {1382-4090},
 volume = {41},
 number = {1-3},
 pages = {233--267},
 year = {2016},
 language = {English},
 doi = {10.1007/s11139-014-9644-x},
 zbMATH = {6664578},
 Zbl = {1418.11064}
}

@misc{franc2018constructionsvectorvaluedmodularforms,
      title={Constructions of vector-valued modular forms of rank four and level one}, 
      author={Cameron Franc and Geoff Mason},
      year={2018},
      eprint={1810.09408},
      archivePrefix={arXiv},
      primaryClass={math.NT},
      url={https://arxiv.org/abs/1810.09408}, 
}

@misc{franc2015structuremodulesvectorvalued,
      title={On the structure of modules of vector valued modular forms}, 
      author={Cameron Franc and Geoffrey Mason},
      year={2015},
      eprint={1509.07494},
      archivePrefix={arXiv},
      primaryClass={math.NT},
      url={https://arxiv.org/abs/1509.07494}, 
}

@article{Marks_2010,
   title={Structure of the module of vector-valued modular forms},
   volume={82},
   ISSN={0024-6107},
   url={http://dx.doi.org/10.1112/jlms/jdq020},
   DOI={10.1112/jlms/jdq020},
   number={1},
   journal={Journal of the London Mathematical Society},
   publisher={Wiley},
   author={Marks, Christopher and Mason, Geoffrey},
   year={2010},
   month=May, pages={32–48} }

@article{brauer1964note,
  author    = {Brauer, Richard},
  title     = {A note on theorems of {Burnside} and {Blichfeldt}},
  journal   = {Proceedings of the American Mathematical Society},
  volume    = {15},
  number    = {1},
  pages     = {31--34},
  year      = {1964},
  publisher = {American Mathematical Society},
  doi       = {10.1007/s0002-9939-1964-0158004-0}
}

@book{burnside1955theory,
  author    = {Burnside, William},
  title     = {Theory of Groups of Finite Order},
  publisher = {Dover Publications, Inc.},
  address   = {New York},
  year      = {1955},
  edition   = {2nd},
  note      = {Unabridged republication of the second edition published in 1911}
}

@misc{steinberg2014burnsidebrauersteinberg,
  author        = {Steinberg, Benjamin},
  title         = {On the {Burnside}-{Brauer}-{Steinberg} theorem},
  year          = {2014},
  eprint        = {1409.7632},
  archivePrefix = {arXiv},
  primaryClass  = {math.RT},
  version       = {2},
  date          = {2014-10-04}
}

@article{Liu:2026ati,
    author = "Liu, Xiang-Gan and Ratz, Michael and Stewart, Alexander",
    title = "{Strong $\mathcal{CP}$ and Quark Mass Hierarchies from Modular Invariance}",
    eprint = "2608.13796",
    archivePrefix = "arXiv",
    primaryClass = "hep-ph",
    month = "8",
    year = "2026"
}

@article{Feruglio:2023uof,
    author = "Feruglio, Ferruccio and Strumia, Alessandro and Titov, Arsenii",
    title = "{Modular invariance and the QCD angle}",
    eprint = "2305.08908",
    archivePrefix = "arXiv",
    primaryClass = "hep-ph",
    doi = "10.1007/JHEP07(2023)027",
    journal = "JHEP",
    volume = "07",
    pages = "027",
    year = "2023"
}

@article{Chen:2019ewa,
    author = "Chen, Mu-Chun and Ramos-S{\'a}nchez, Sa{\'u}l and Ratz, Michael",
    title = "{A note on the predictions of models with modular flavor symmetries}",
    eprint = "1909.06910",
    archivePrefix = "arXiv",
    primaryClass = "hep-ph",
    reportNumber = "UCI-TR-2019-22, TUM-HEP 1223/19",
    doi = "10.1016/j.physletb.2019.135153",
    journal = "Phys. Lett. B",
    volume = "801",
    pages = "135153",
    year = "2020"
}

@article{Baur:2019kwi,
    author = "Baur, Alexander and Nilles, Hans Peter and Trautner, Andreas and Vaudrevange, Patrick K. S.",
    title = "{Unification of Flavor, CP, and Modular Symmetries}",
    eprint = "1901.03251",
    archivePrefix = "arXiv",
    primaryClass = "hep-th",
    reportNumber = "LMU-ASC 02/19, TUM-HEP 1181/19",
    doi = "10.1016/j.physletb.2019.03.066",
    journal = "Phys. Lett. B",
    volume = "795",
    pages = "7--14",
    year = "2019"
}

@article{Nilles:2020nnc,
    author = "Nilles, Hans Peter and Ramos-S{\'a}nchez, Sa{\'u}l and Vaudrevange, Patrick K. S.",
    title = "{Eclectic Flavor Groups}",
    eprint = "2001.01736",
    archivePrefix = "arXiv",
    primaryClass = "hep-ph",
    reportNumber = "TUM-HEP 1247/19",
    doi = "10.1007/JHEP02(2020)045",
    journal = "JHEP",
    volume = "02",
    pages = "045",
    year = "2020"
}

@article{Nilles:2020kgo,
    author = "Nilles, Hans Peter and Ramos-Sanchez, Saul and Vaudrevange, Patrick K. S.",
    title = "{Lessons from eclectic flavor symmetries}",
    eprint = "2004.05200",
    archivePrefix = "arXiv",
    primaryClass = "hep-ph",
    reportNumber = "TUM-HEP 1256/20",
    doi = "10.1016/j.nuclphysb.2020.115098",
    journal = "Nucl. Phys. B",
    volume = "957",
    pages = "115098",
    year = "2020"
}

@article{Chen:2021prl,
    author = "Chen, Mu-Chun and Knapp-Perez, Victor and Ramos-Hamud, Mario and Ramos-Sanchez, Saul and Ratz, Michael and Shukla, Shreya",
    title = "{Quasi{\textendash}eclectic modular flavor symmetries}",
    eprint = "2108.02240",
    archivePrefix = "arXiv",
    primaryClass = "hep-ph",
    reportNumber = "UCI-TR-2021-20",
    doi = "10.1016/j.physletb.2021.136843",
    journal = "Phys. Lett. B",
    volume = "824",
    pages = "136843",
    year = "2022"
}

@article{Baur:2022hma,
    author = "Baur, Alexander and Nilles, Hans Peter and Ramos-Sanchez, Saul and Trautner, Andreas and Vaudrevange, Patrick K. S.",
    title = "{The first string-derived eclectic flavor model with realistic phenomenology}",
    eprint = "2207.10677",
    archivePrefix = "arXiv",
    primaryClass = "hep-ph",
    reportNumber = "TUM-HEP 1408/22",
    doi = "10.1007/JHEP09(2022)224",
    journal = "JHEP",
    volume = "09",
    pages = "224",
    year = "2022"
}

@article{Li:2024pff,
    author = "Li, Cai-Chang and Lu, Jun-Nan and Ding, Gui-Jun",
    title = "{Minimal eclectic flavor group Q$_{8}$ {\ensuremath{\rtimes}} S$_{3}$ and neutrino mixing}",
    eprint = "2405.13460",
    archivePrefix = "arXiv",
    primaryClass = "hep-ph",
    doi = "10.1007/JHEP12(2024)015",
    journal = "JHEP",
    volume = "12",
    pages = "015",
    year = "2024"
}

@article{Li:2023dvm,
    author = "Li, Cai-Chang and Ding, Gui-Jun",
    title = "{Eclectic flavor group~$\Delta(27)\rtimes S_3$ and lepton model building}",
    eprint = "2308.16901",
    archivePrefix = "arXiv",
    primaryClass = "hep-ph",
    doi = "10.1007/JHEP03(2024)054",
    journal = "JHEP",
    volume = "03",
    pages = "054",
    year = "2024"
}

@article{Feruglio:2026xxo,
    author = "Feruglio, Ferruccio and Marrone, Antonio",
    title = "{Eclectic flavour symmetries without flavons}",
    eprint = "2609.16118",
    archivePrefix = "arXiv",
    primaryClass = "hep-ph",
    month = "9",
    year = "2026"
}

@article{Ding:2020zxw,
    author = "Ding, Gui-Jun and Feruglio, Ferruccio and Liu, Xiang-Gan",
    title = "{Automorphic Forms and Fermion Masses}",
    eprint = "2010.07952",
    archivePrefix = "arXiv",
    primaryClass = "hep-th",
    reportNumber = "USTC-ICTS/PCFT-20-28",
    doi = "10.1007/JHEP01(2021)037",
    journal = "JHEP",
    volume = "01",
    pages = "037",
    year = "2021"
}

@article{Carducci:2026shs,
    author = "Carducci, M. and Meloni, D. and Parriciatu, M. and Penedo, J. T.",
    title = "{Quark hierarchies and CP violation from the Siegel modular group}",
    eprint = "2604.21979",
    archivePrefix = "arXiv",
    primaryClass = "hep-ph",
    doi = "10.1007/JHEP09(2026)268",
    journal = "JHEP",
    volume = "09",
    pages = "268",
    year = "2026"
}

@article{deMedeirosVarzielas:2019cyj,
    author = "de Medeiros Varzielas, Ivo and King, Stephen F. and Zhou, Ye-Ling",
    title = "{Multiple modular symmetries as the origin of flavor}",
    eprint = "1906.02208",
    archivePrefix = "arXiv",
    primaryClass = "hep-ph",
    doi = "10.1103/PhysRevD.101.055033",
    journal = "Phys. Rev. D",
    volume = "101",
    number = "5",
    pages = "055033",
    year = "2020"
}

@article{Qu:2024rns,
    author = "Qu, Bu-Yao and Ding, Gui-Jun",
    title = "{Non-holomorphic modular flavor symmetry}",
    eprint = "2406.02527",
    archivePrefix = "arXiv",
    primaryClass = "hep-ph",
    doi = "10.1007/JHEP08(2024)136",
    journal = "JHEP",
    volume = "08",
    pages = "136",
    year = "2024"
}

@article{Qu:2025ddz,
    author = "Qu, Bu-Yao and Lu, Jun-Nan and Ding, Gui-Jun",
    title = "{Non-holomorphic modular flavor symmetry and odd weight polyharmonic Maa{\ss} form}",
    eprint = "2506.19822",
    archivePrefix = "arXiv",
    primaryClass = "hep-ph",
    doi = "10.1007/JHEP11(2025)140",
    journal = "JHEP",
    volume = "11",
    pages = "140",
    year = "2025"
}

@article{Ding:2024inn,
    author = "Ding, Gui-Jun and Lu, Jun-Nan and Petcov, S. T. and Qu, Bu-Yao",
    title = "{Non-holomorphic modular S$_{4}$ lepton flavour models}",
    eprint = "2408.15988",
    archivePrefix = "arXiv",
    primaryClass = "hep-ph",
    reportNumber = "SISSA 17/2024/FISI",
    doi = "10.1007/JHEP01(2025)191",
    journal = "JHEP",
    volume = "01",
    pages = "191",
    year = "2025"
}

@article{Petcov:2023fwh,
    author = "Petcov, S. T.",
    title = "{On the normalisation of the modular forms in modular invariant theories of flavour}",
    eprint = "2311.04185",
    archivePrefix = "arXiv",
    primaryClass = "hep-ph",
    reportNumber = "SISSA 14/2023/FISI",
    doi = "10.1016/j.physletb.2024.138540",
    journal = "Phys. Lett. B",
    volume = "850",
    pages = "138540",
    year = "2024"
}

@misc{nufit52,
    author        = "{NuFIT Collaboration}",
    title         = "{NuFIT 5.2 (2022): Three-neutrino fit based on data available in November 2022}",
    howpublished  = "\url{http://www.nu-fit.org/?q=node/256}",
    year          = "2022"
}

@article{Esteban:2020cvm,
    author = "Esteban, Ivan and Gonzalez-Garcia, M. C. and Maltoni, Michele and Schwetz, Thomas and Zhou, Albert",
    title = "{The fate of hints: updated global analysis of three-flavor neutrino oscillations}",
    eprint = "2007.14792",
    archivePrefix = "arXiv",
    primaryClass = "hep-ph",
    reportNumber = "IFT-UAM/CSIC-112, YITP-SB-2020-21",
    doi = "10.1007/JHEP09(2020)178",
    journal = "JHEP",
    volume = "09",
    pages = "178",
    year = "2020"
}

@article{Antusch:2013jca,
    author = "Antusch, Stefan and Maurer, Vinzenz",
    title = "{Running quark and lepton parameters at various scales}",
    eprint = "1306.6879",
    archivePrefix = "arXiv",
    primaryClass = "hep-ph",
    doi = "10.1007/JHEP11(2013)115",
    journal = "JHEP",
    volume = "11",
    pages = "115",
    year = "2013"
}

@manual{GAP4102,
  title        = {{GAP -- Groups, Algorithms, and Programming, Version 4.10.2}},
  organization = {The GAP Group},
  year         = {2019},
  url          = {https://www.gap-system.org},
}

@misc{Victor,
      year={to appear},
      author={Bonnefoy, Quentin and Rausch de Traubenberg, Michel and Saulquin, Victor}}

@misc{GitHub_vsaulquin,
      howpublished={\url{https://github.com/vsaulquin/Algorithmic-construction-of-VVMFs}}}

\end{document}